\documentclass[useAMS,usenatbib]{mn2e}
\usepackage{amssymb}
\usepackage[dvips]{graphicx}
\usepackage{natbib}

\title[Magnetar powered GRBs]{Magnetar powered GRBs: Explaining the extended emission and X-ray plateau of short GRB light curves}
\author[B. P. Gompertz, P. T. O'Brien and G. A. Wynn]{B. P. Gompertz\thanks{E-mail: bpg6@le.ac.uk}, P. T. O'Brien, and G. A. Wynn\\
Department of Physics and Astronomy, University of Leicester, Leicester, UK, LE1 7RH}

\begin{document}

\date{Accepted:}

\pagerange{\pageref{firstpage}--\pageref{lastpage}} \pubyear{????}

\maketitle

\label{firstpage}

\begin{abstract}
Extended emission (EE) is a high-energy, early time rebrightening sometimes seen in the light curves of short gamma-ray bursts (GRBs). We present the first contiguous fits to the EE tail and the later X-ray plateau, unified within a single model. Our central engine is a magnetar surrounded by a fall-back accretion disc, formed by either the merger of two compact objects or the accretion-induced collapse of a white dwarf. During the EE phase, material is accelerated to super-Keplarian velocities and ejected from the system by the rapidly rotating ($P \approx 1 - 10$ ms) and very strong ($10^{15}$ G) magnetic field in a process known as magnetic propellering. The X-ray plateau is modelled as magnetic dipole spin-down emission. We first explore the range of GRB phenomena that the propeller could potentially reproduce, using a series of template light curves to devise a classification scheme based on phenomology. We then obtain fits to the light curves of 9 GRBs with EE, simultaneously fitting both the propeller and the magnetic dipole spin-down and finding typical disc masses of a few $10^{-3}$ $M_{\odot}$ to a few $10^{-2}$ $M_{\odot}$. This is done for ballistic, viscous disc and exponential accretion rates. We find that the conversion efficiency from kinetic energy to EM emission for propellered material needs to be $\gtrsim 10\%$ and that the best fitting results come from an exponential accretion profile.
\end{abstract}

\section{Introduction}

Magnetars are a subset of neutron stars (NS) with extremely high magnetic fields that can exceed $10^{15}$ G at birth \citep{Duncan92}. These dipole fields can be achieved through a number of different processes, eg an $\alpha$ -- $\Omega$ dynamo \citep{Duncan92}, shear instabilities during compact object merger \citep{Price06}, or magneto-rotational instabilities during core collapse (MRI; \citealt{Akiyama03,Thompson05}). Evidence for the existence of magnetars comes from observations of soft gamma-ray repeaters (SGR; \citealt{Norris91}). The relative hardness and extreme luminosities of these events suggest they identify with NSs with dipole fields of around $10^{15}$ G \citep{Thompson95}, despite being millions of years old. \citet{Thompson95} also present six independent arguments for a birth dipole field of order $\sim 10^{15}$ G for the magnetar behind the $10^{44}$ erg March 5 event \citep{Mazets79} seen in SGR 0526 -- 66 in the $\sim 10^{4}$ yr old supernova remnant N49. A number of other SGR events have been studied, and the central engines confirmed to be magnetars with strong ($\sim 10^{14}$ -- $10^{15}$ G) dipole magnetic fields (eg \citealt{Kouveliotou98,Kouveliotou99,Woods99,Esposito10}).

The birth and early-time spin-down of these magnetars has been suggested as a potential progenitor for gamma-ray bursts (GRBs), both long (LGRB; \citealt{Zhang01,Lyons10,Dall'Osso11,Metzger11,Bernardini12}) and short (SGRB; \citealt{Fan06,Rowlinson10,Rowlinson13}). These classes are based on a temporal bimodality \citep{Kouveliotou93} seen in the parameter $T_{90}$; the time in which 90\% of the gamma-ray fluence is detected. In theory, LGRBs have $T_{90}$ $>$ 2 seconds and SGRBs have $T_{90}$ $<$ 2 seconds, but this partition is not absolute (eg \citealt{Gehrels06,Kann11}). In the case of SGRBs, the magnetar is said to be formed by the merger of two compact objects, either a NS binary, a white dwarf (WD) binary, or a NS -- WD binary \citep{Paczynski86,Fryer99,Rosswog03,Belczynski06,Chapman07,Giacomazzo13b} or the accretion-induced collapse of a WD \citep{Metzger08}. The alternative to the merger scenario is the collapsar model \citep{Woosley93,Paczynski98,MacFadyen99}, which is favoured amongst LGRBs due to the accompaniment of a type Ib/c supernova in every case where it would be possible to detect one \citep{Galama98,Stanek03}. These supernova signatures have never been observed in SGRBs (eg \citealt{Kann11}).

Within the SGRB class, there is also a subsection of bursts that exhibit a rebrightening in their high energy light curves after the initial emission spike, at times of around $10$ s after trigger. These are known as extended emission (EE) bursts \citep{Norris06}. EE typically has lower peak flux than the initial spike, but can last up to a few hundred seconds, meaning that the total fluence of EE is often higher \citep{Perley09}. Finding a central engine that can recreate this emission feature has proved to be a problem, although a number of different models have been proposed \citep{Rosswog07,Metzger10,Barkov11,Zhang11}. In the present work, we explore the possibility that EE is the product of an accretion disc around a magnetar undergoing magnetic propellering. Magnetar spin-down is not a new idea as the central engine behind EE GRBs \citep{Metzger08,Bucciantini12,Gompertz13}, and magnetar winds have previously been suggested as the source of EE by \citet{Metzger11} and \citet{Bucciantini12}, but magnetic propellering has so far not been directly tested against the light curves as the energy extraction mechanism (however see \citealt{Piro11} and \citealt{Bernardini13} for its application on LGRBs).

The mechanisms employed for each light curve feature are discussed in Section 2, and model parameter space is explored in section 3. This model is then fitted to the data in section 4 and the findings are discussed in section 5. The main conclusions are summarised in section 6.

\section{Emission mechanics}

\subsection{Prompt emission}
Within the framework of the compact object binary merger, prompt emission is often said to be the accretion of a disc or torus onto the newly formed protomagnetar \citep{Narayan01,Metzger08,Metzger10,Bucciantini12}. We will assume the same here, and concern ourselves more with the mechanics behind the EE tail and the late-time plateau. As the compact objects spiral inwards, simulations suggest that some material (possibly up to $10^{-1}$ M$_{\odot}$, \citealt{Lee09}) is ejected by tidal disruption into a tidal tail through the outer Lagrange point. \citet{Lee09} find that this material returns at $\sim 1 - 10$ s, and creates a new ring at a radius of around $300 - 500$ km, with a mass $M_{fb} \approx 10^{-2}$ M$_{\odot}$. Similar behaviour was found by \citet{Rosswog07}, who showed that the range of fallback behaviours is much more varied in an unequal mass binary. For the formation of a magnetar, this would mean a NS -- WD system, or a NS binary involving a more massive NS (see e.g. \citealt{Demorest10}). The result is that after the torus is accreted and prompt emission has been produced, we are left with a rapidly-rotating magnetar surrounded by a $\sim 10^{-6} - 10^{-1}$ M$_{\odot}$ accretion disc at a radius of a few hundred km.

\subsection{Extended emission}
\begin{figure}
\begin{center}
\begin{tabular}{ l p{5cm} }
\raisebox{-\totalheight}{\includegraphics[width=2.8cm,angle=-0]{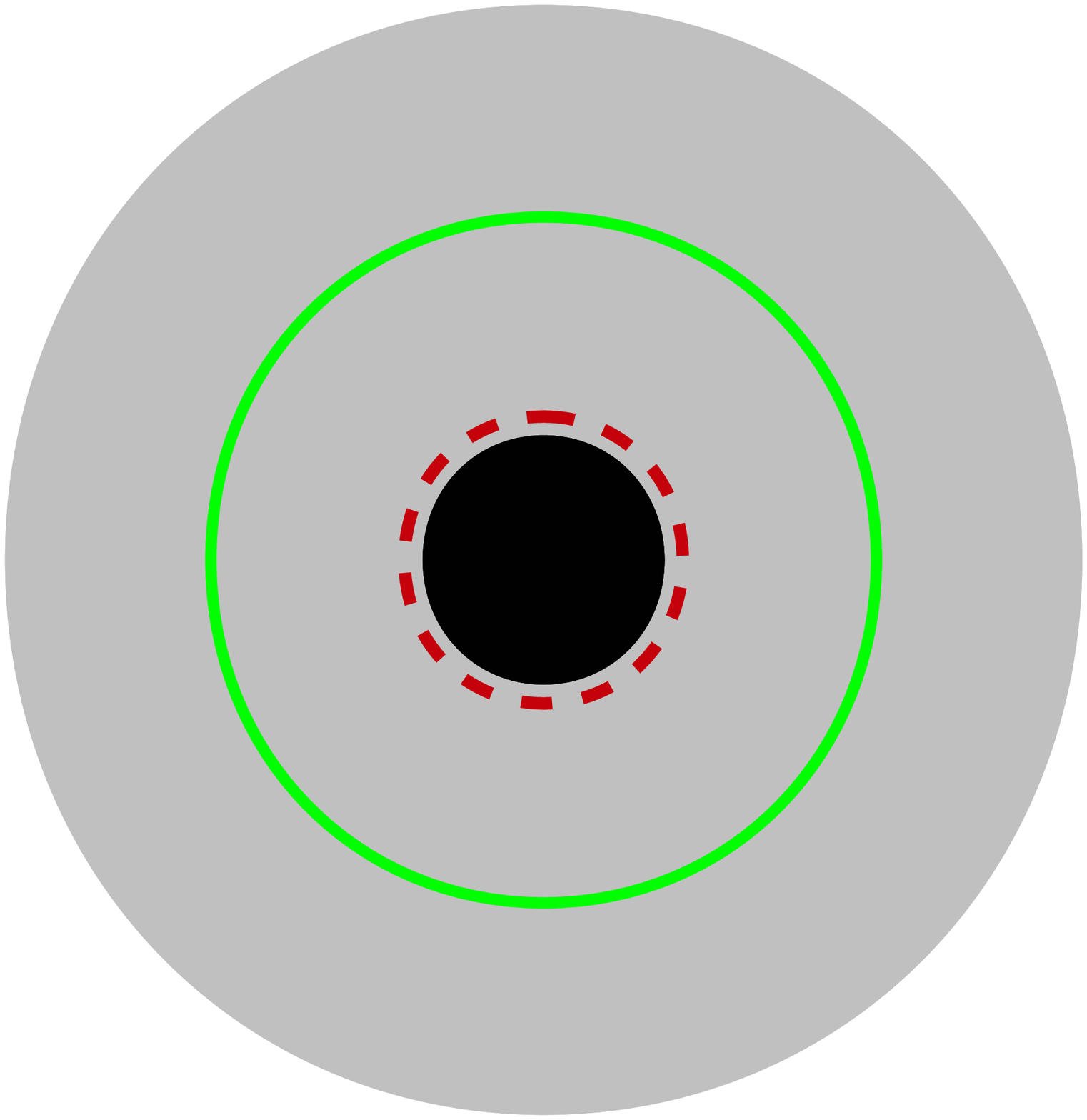}} & \textbf{a)} During the early stages of accretion, the Alfv\'{e}n radius may be suppressed by a high $\dot{M}$ so that $r_c > r_m$ and the magnetar is spun up by accretion. The increased spin period will cause $r_c$ to shrink. However, if initial accretion is not sufficiently high, the system will begin propellering right away. \\
 & \\
\raisebox{-\totalheight}{\includegraphics[width=2.8cm,angle=-0]{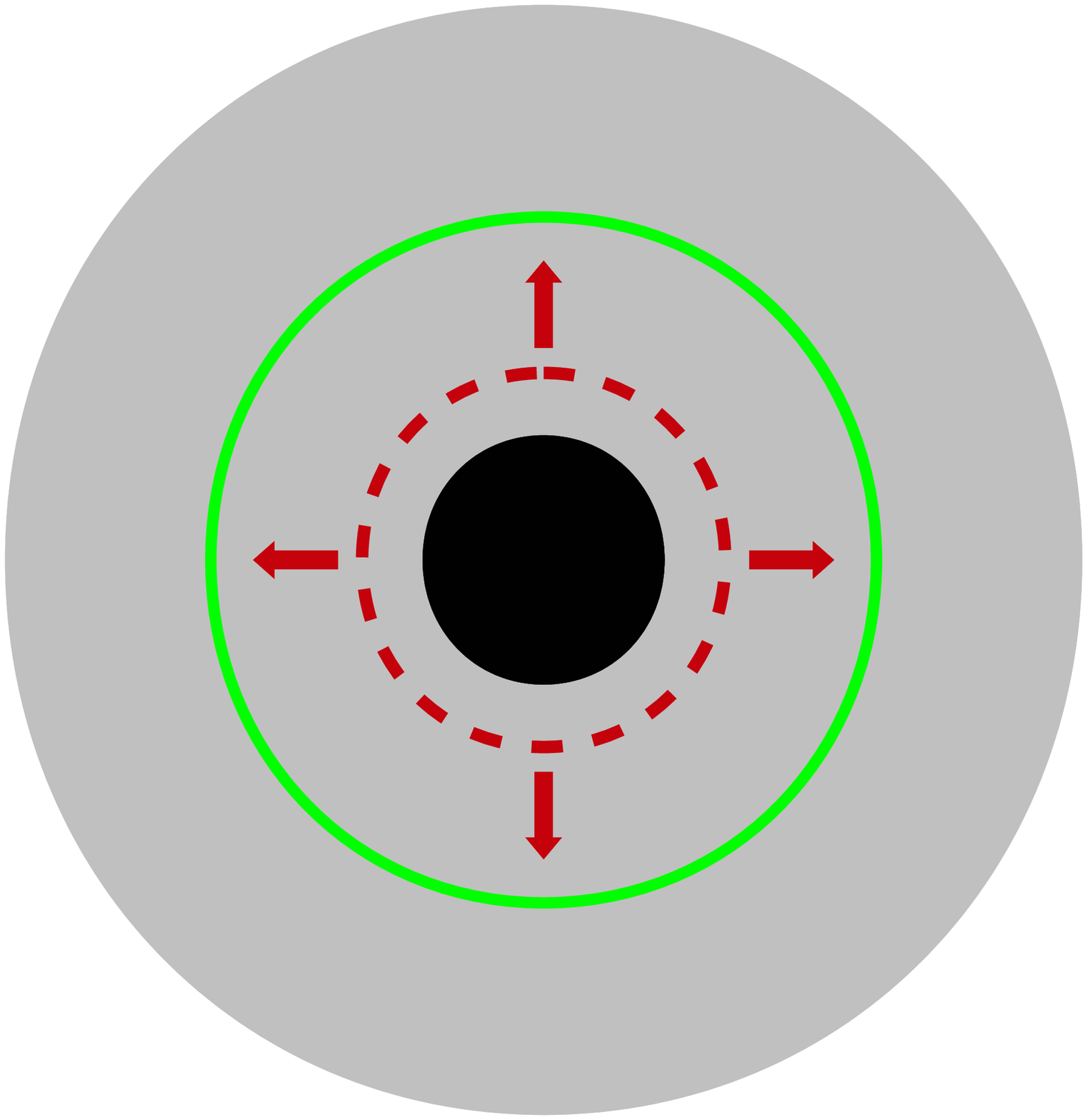}} & \textbf{b)} As accretion falls off (see equation 13), the Alfv\'{e}n radius expands. \\
 & \\
\raisebox{-\totalheight}{\includegraphics[width=2.8cm,angle=-0]{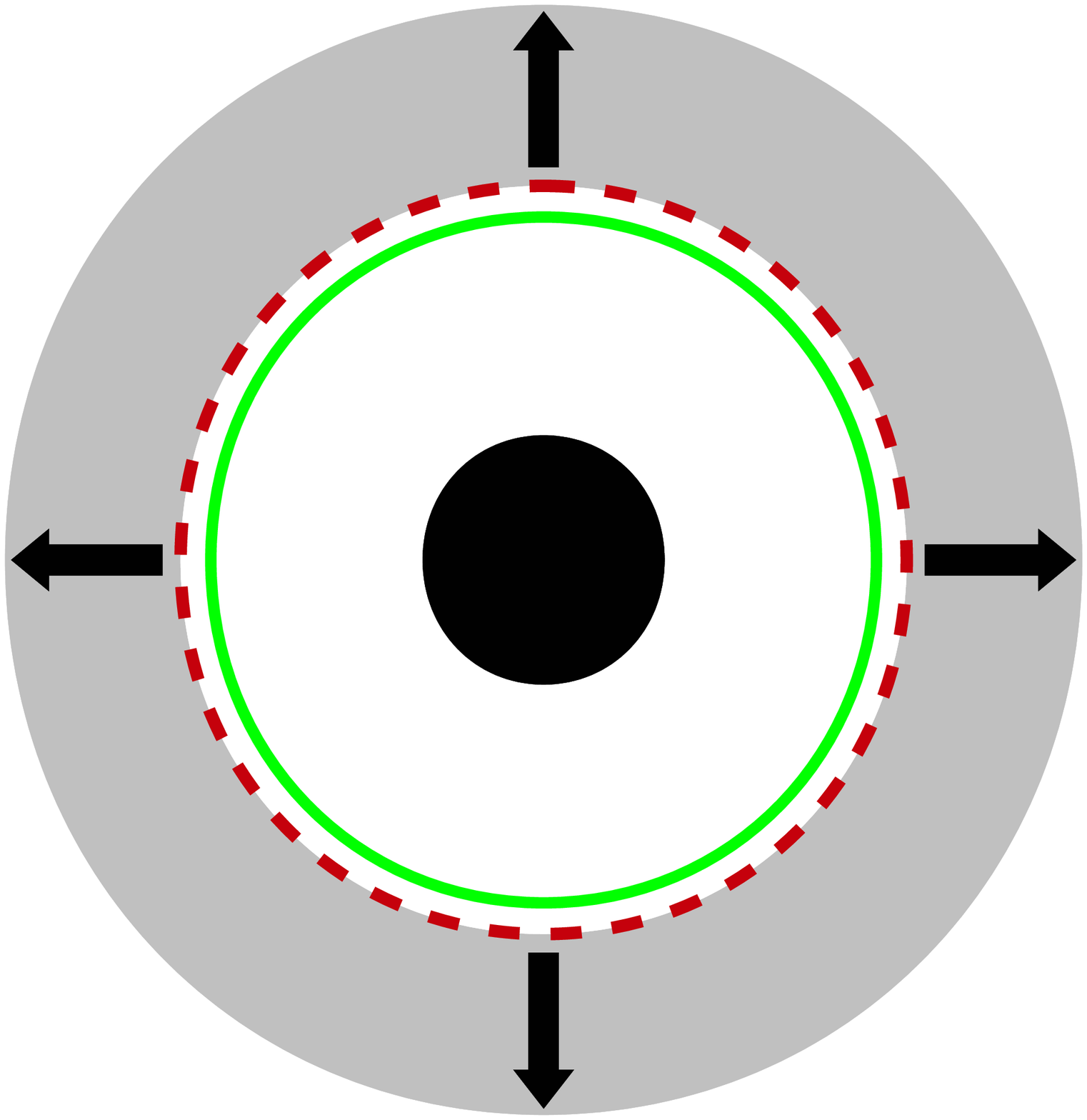}} & \textbf{c)} Once $r_m > r_c$ the system enters the propeller regime. Material already within $r_c$ accretes on to the surface of the magnetar, whilst material falling in from greater radii is propellered away at $r_m$. If the propeller is not strong enough for material to escape the potential well, no emission is seen and material returns to the disc. \\
 & \\
\raisebox{-\totalheight}{\includegraphics[width=2.8cm,angle=-0]{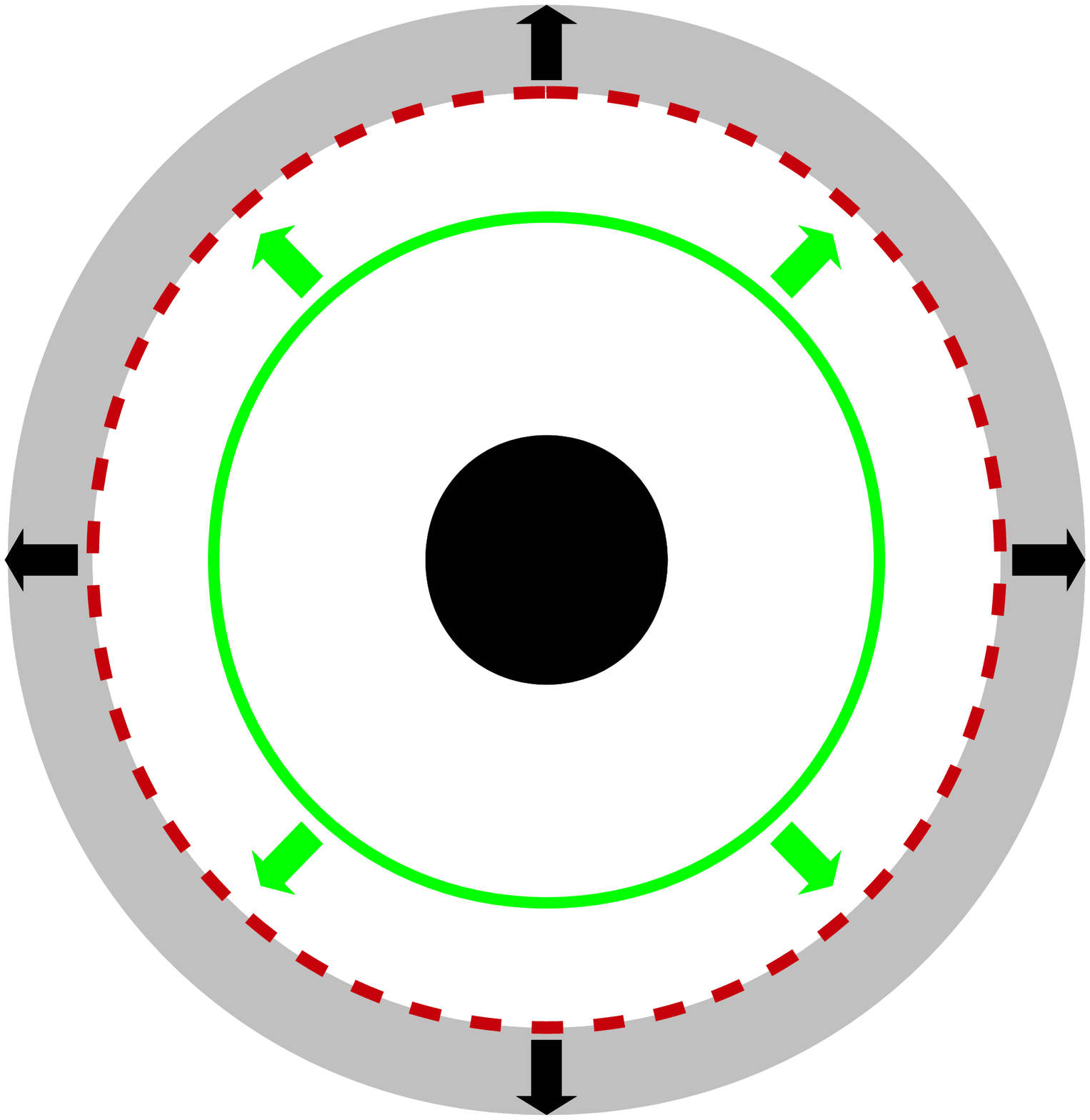}} & \textbf{d)} $r_m$ continues to expand as the accretion rate drops, but the loss of angular momentum to the expelled material means the magnetar begins to spin more slowly, causing the expansion of $r_c$. If $r_c$ outgrows $r_m$, the system will begin to accrete again. \\
 & \\
\raisebox{-\totalheight}{\includegraphics[width=2.8cm,angle=-0]{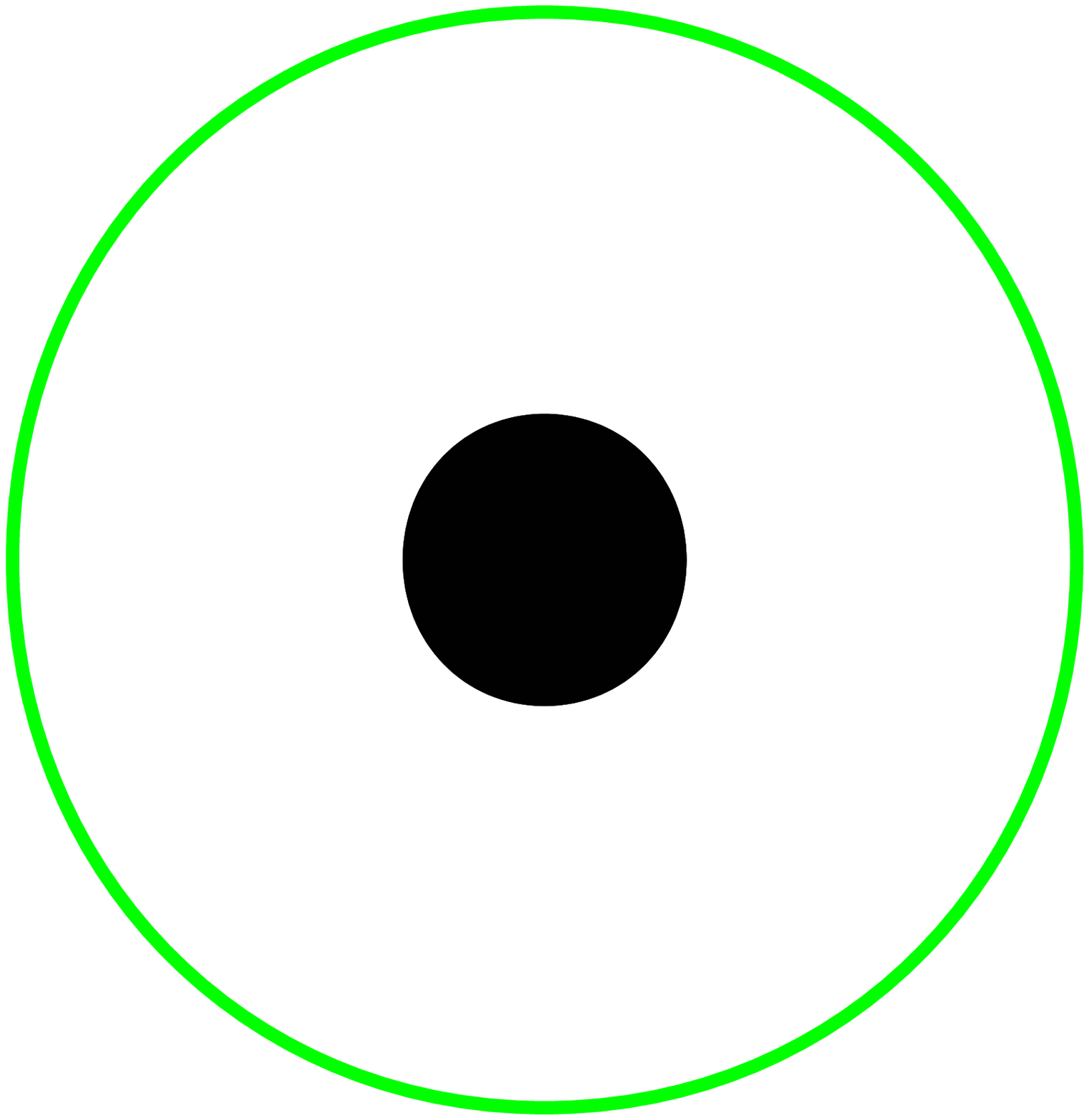}} & \textbf{e)} When the accretion disc is depleted, $r_m$ becomes essentially infinite and plays no further part. The new value of $r_c$ is set by the spin period of the magnetar, and slowly increases as spin is lost to dipole emission. \\
\end{tabular}
\caption{A toy model describing the interaction of the Alfv\'{e}n radius ($r_m$) and the co-rotation radius ($r_c$) during the propellering and accretion regimes. The black circle is the central magnetar. The grey region represents the accretion disc. The red dashed line indicates the Alfv\'{e}n radius, whilst the green solid line denotes the co-rotation radius. Not all stages may be present in an individual burst. Some may occur twice.}
\label{propplot}
\end{center}
\end{figure}

The model used for EE in the present work is the magnetic propeller model of \citet{Piro11} and is summarised in Figure~\ref{propplot}. The magnetic dipole field of the central magnetar is given by $B = \mu /r^3$, where $\mu = B_* R^3$ is the magnetic dipole moment for a star with surface dipole field $B_*$ and radius $R$. The magnetic pressure for a given radius, $r$, is then
\begin{equation}
P_{mag}= \frac{\mu^2}{2 \mu_0 r^6}
\end{equation}
Material falling in from the accretion disc also exerts its own force, opposing that of $P_{mag}$. This is the ram pressure, given by
\begin{equation}
P_{ram}= \frac{\dot{M}}{8\pi}\bigg{(}\frac{2GM_*}{r^5}\bigg{)}^{1/2}
\end{equation}
where $M_*$ is the mass of the magnetar. Equating these two pressures gives the radius at which infalling material comes under strong influence from the dipole field, known as the Alfv\'{e}n radius, $r_m$.
\begin{equation}
r_m = \mu^{4/7}(GM_*)^{-1/7}\dot{M}^{-2/7}
\end{equation}
This is one of the two key radii that determine the behaviour of the magnetar, the other being the corotation radius, $r_c$, the radius at which material orbits at the same rate as the stellar surface. 
\begin{equation}
r_c = (GM_*/\Omega^2)^{1/3}
\end{equation}
where $\Omega = \frac{2\pi}{P}$ is the angular frequency of the magnetar and $P$ is the spin period. If $r_c > r_m$, the accretion disc is rotating more rapidly than the magnetic field at the point the field becomes dynamically important, so the effect of the interaction is to slow the material and allow it to accrete (Figure~\ref{propplot}a and~\ref{propplot}b). The accreting material also spins up the magnetar, and therefore the field. If $r_c < r_m$ however, the magnetic field is spinning faster than the material, and the interaction causes particles to be accelerated to super-Keplerian velocities and ejected from the system. The magnetar loses angular momentum to the expelled material via the magnetic field and is slowed. This condition, with $r_c < r_m$, is the propeller regime (Figure~\ref{propplot}c and~\ref{propplot}d). Since material cannot be accelerated to the speed of light (or above), $r_m$ must be capped at some realistic fraction, $k$, of the light cylinder radius, $r_{lc}$. This radius marks the point at which the magnetic field lines must orbit at the speed of light to maintain their rigid rotation with the stellar surface, and is defined as

\begin{equation}
r_{lc} = c/\Omega
\end{equation}
The value of $k$ naturally sets the maximum particle ejection velocity as $v=kc$.

These two regimes, propeller and accretion, both affect the spin period of the central magnetar. If $r_m > R$, the accretion torque, $N_{acc}$, is given by
\begin{equation}
N_{acc} = n(\omega)(GM_*r_m)^{1/2}\dot{M}
\end{equation}
$n(\omega)$ is the dimensionless torque, where the `fastness parameter,' $\omega = \Omega/(GM_*/r_m^3)^{1/2} = (r_m/r_c)^{3/2}$ and $n = 1 - \omega$. If $r_m < R$, the torque becomes
\begin{equation}
N_{acc} = (1-\Omega/\Omega_K)(GM_*R)^{1/2}\dot{M}
\end{equation}
where $\Omega_K = (GM_*/R^3)^{1/2}$. The accretion torque will spin up the magnetar when $r_m < r_c$, but goes negative in cases where $r_m > r_c$ to account for the angular momentum lost with propellered material. The other contribution to the torque comes from dipole spin-down, $N_{dip}$, and is given by
\begin{equation}
N_{dip} = - \frac{2}{3}\frac{\mu^2\Omega^3}{c^3}\bigg{(}\frac{r_{lc}}{r_m}\bigg{)}^3
\end{equation}
Equation 8 takes into account the enhanced dipole spin-down that results from the additional open field lines created by an accretion disc truncating the magnetosphere at a radius less than $r_{lc}$, and is taken from equation 2 of \citet{Bucciantini06}, who give a good discussion of this point. From these two contributions, the change in spin can be calculated by
\begin{equation}
\dot{\Omega}=\frac{N_{dip} + N_{acc}}{I}
\end{equation}
where $I = 0.35M_*R^2$ is the moment of inertia. As the spin changes, we must keep track of the rotation parameter, $\beta \equiv T/ \vert W \vert$, where $T = \frac{1}{2}I\Omega^2$ is the rotational energy and $\vert W \vert$ is the binding energy. We follow \citet{Piro11} in using the prescription from \citet{Lattimer01} for $\vert W \vert$,
\begin{equation}
\vert W \vert \approx 0.6M_*c^2\frac{GM_*/Rc^2}{1 - 0.5(GM_*/Rc^2)}
\end{equation}
$R$ is kept constant, even if $M_*$ is increased by accretion, since this is consistent with most equations of state \citep{Lattimer01}. If $\beta > 0.27$, dynamical bar-mode instability will radiate or hydrodynamically readjust angular momentum, so we set $N_{acc} = 0$ when $\beta > 0.27$. Collecting all these terms together, we estimate the kinetic luminosity of the propeller material as
\begin{equation}
L_{prop} = -N_{acc}\Omega - GM_*\dot{M}/r_m
\end{equation}
The first term is the emission luminosity, and is negative because $N_{acc}$ has been defined as negative when the magnetar is spinning down. The second term represents the energy required to escape from the gravitational potential well. This equation implicitly assumes material outflow originates from the inner edge of the disc. It therefore represents a lower limit for kinetic luminosity, as material escaping from further out will lose less energy in doing so.

We assume a thick disc, with scale height, $H$, equal to the \textbf{outer} disc radius $R_d$. Fallback material returns to the disc at the ballistic fallback rate of $t^{-5/3}$, but must shed its angular momentum before accreting onto the central NS. In systems such as these, the accretion rate is commonly modelled as a viscous disc with a $t^{-4/3}$ profile (see e.g. \citealt{Cannizzo11}), however in the presence of strong outflows \citep{Fernandez13}, the accretion rate will proceed as an exponential. We will attempt to model all 3 accretion profiles in an effort to gauge the sensitivity of the results to them. We adopt the form of the initial accretion rate (c.f. \citealt{King98})
\begin{equation}
\dot{M}_0 = M_d3\nu /R_d^2
\end{equation}
where $M_d$ is the initial disc mass and $\nu$ is the viscosity. Accretion then proceeds either as one of the two power laws mentioned above, or as an exponential decay of the form
\begin{equation}
\dot{M} = \dot{M}_0 e^{-3\nu t/R_d^2}
\end{equation}

\subsection{Dipole spin-down}
To explain the late-time plateau ($\sim 10^3$ -- $10^4$ s), we invoke the contribution to the light curve from dipole spin-down, based on the model in \citet{Zhang01}. This has been done previously on LGRBs \citep{Lyons10,Dall'Osso11,Bernardini12}, SGRBs \citep{Fan06,Rowlinson13} and EE GRBs by \citet{Gompertz13}. These works assumed a constant rate of spin-down, and therefore a constant level of dipole luminosity, however during EE the spin period may be highly variable, making this a simplified approximation. Since we are recording the evolution of the spin period in the magnetic propeller model, the time-varying equations \citep{Zhang01} for dipole emission can be used. The luminosity contribution from dipole spin-down is
\begin{equation}
L_{dip} = \mu^2\Omega^4/6c^3
\end{equation}
This emission component can be highly variable during propellering, but will settle to a constant level once the accretion disc has been consumed. As the propeller luminosity fades, $L_{dip}$ will begin to show up in the light curve, causing the flattening seen in the late-time plateau.

\begin{table}
\begin{center}
\begin{tabular}{lllllll}
\hline
$B$ & ($10^{15} G$) & 1 & 5 & 10 & 50 & - \\
$P$ & (ms) & 1 & 5 & 10 & - & - \\
$M_d$ & ($M_{\odot}$) & $10^{-5}$ & $10^{-4}$ & $10^{-3}$ & $10^{-2}$ & $10^{-1}$ \\
$R_d$ & (km) & 100 & 500 & 1000 & - & - \\
$\alpha$ & & 0.1 & 0.2 & 0.3 & 0.4 & 0.5 \\
$c_s$ & ($10^7$cm s$^{-1}$) & 1 & 2 & 3 & - & - \\
$M_*$ & ($M_{\odot}$) & 1.4 & 2.0 & 2.5 & - & - \\
\hline
\end{tabular}
\caption{Values used to test the morphological effects of parameter variation. The total number of combinations resulted in 8100 synthetic light curves. $B$ - Magnetic field; $P$ - Spin period; $M_d$ - Disc mass; $R_d$ - Disc radius; $\alpha$ - Viscosity in the disc; $c_s$ - Sound speed in the disc; $M_*$ - Mass of the central magnetar.}
\end{center}
\end{table}

\begin{figure*}
\begin{center}
\includegraphics[width=7.5cm,angle=-0]{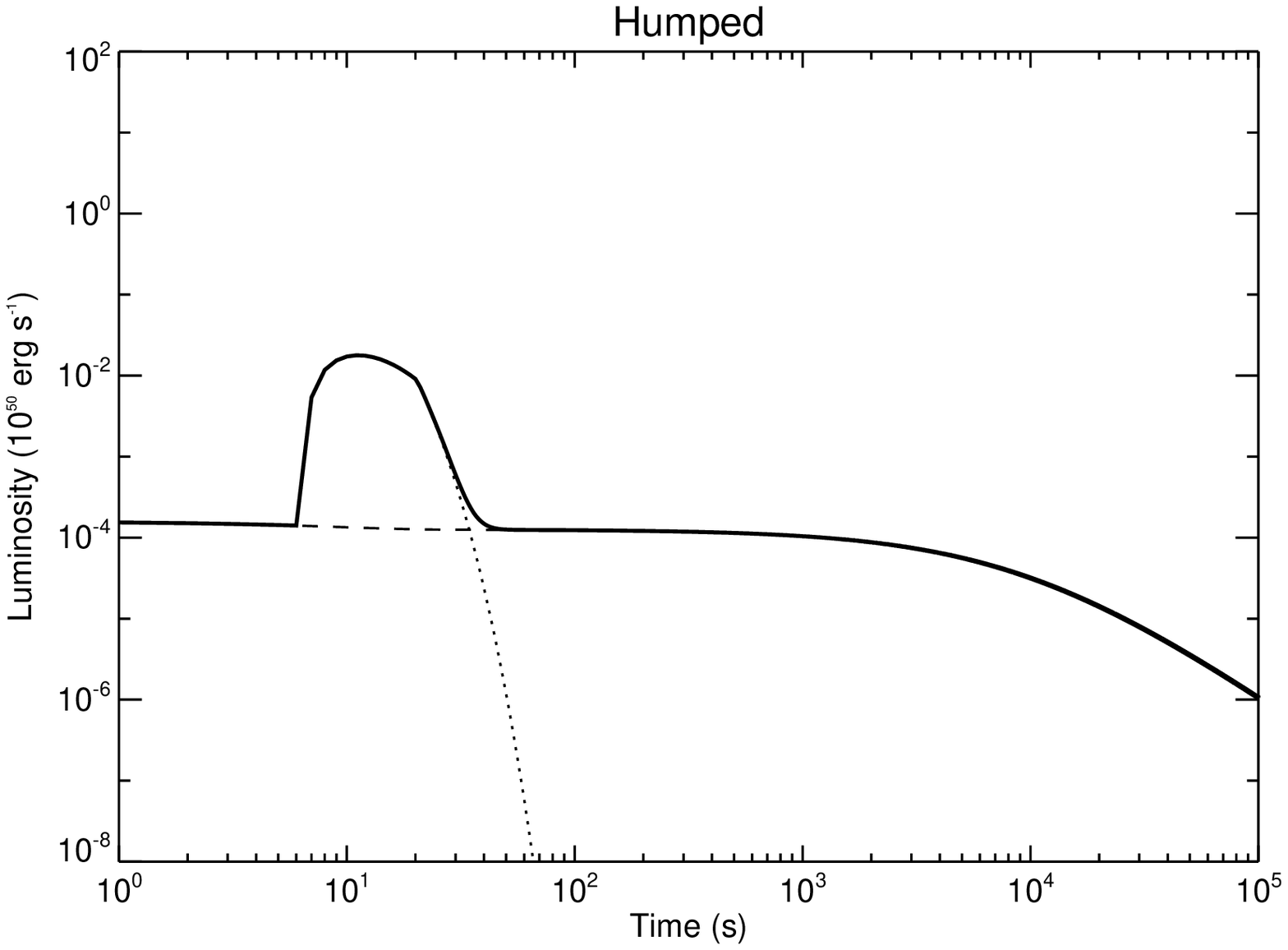}
\includegraphics[width=7.5cm,angle=-0]{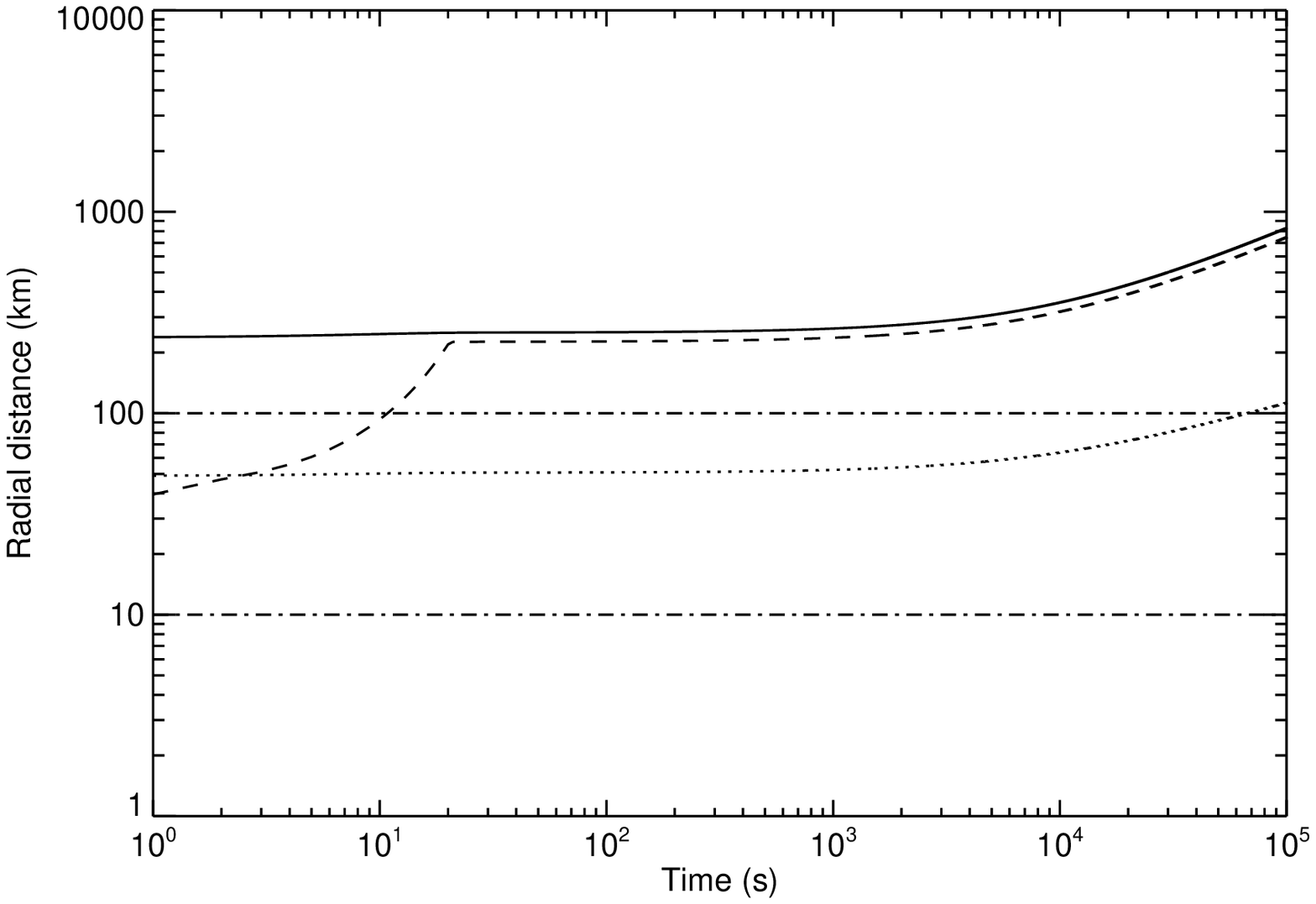}
\includegraphics[width=7.5cm,angle=-0]{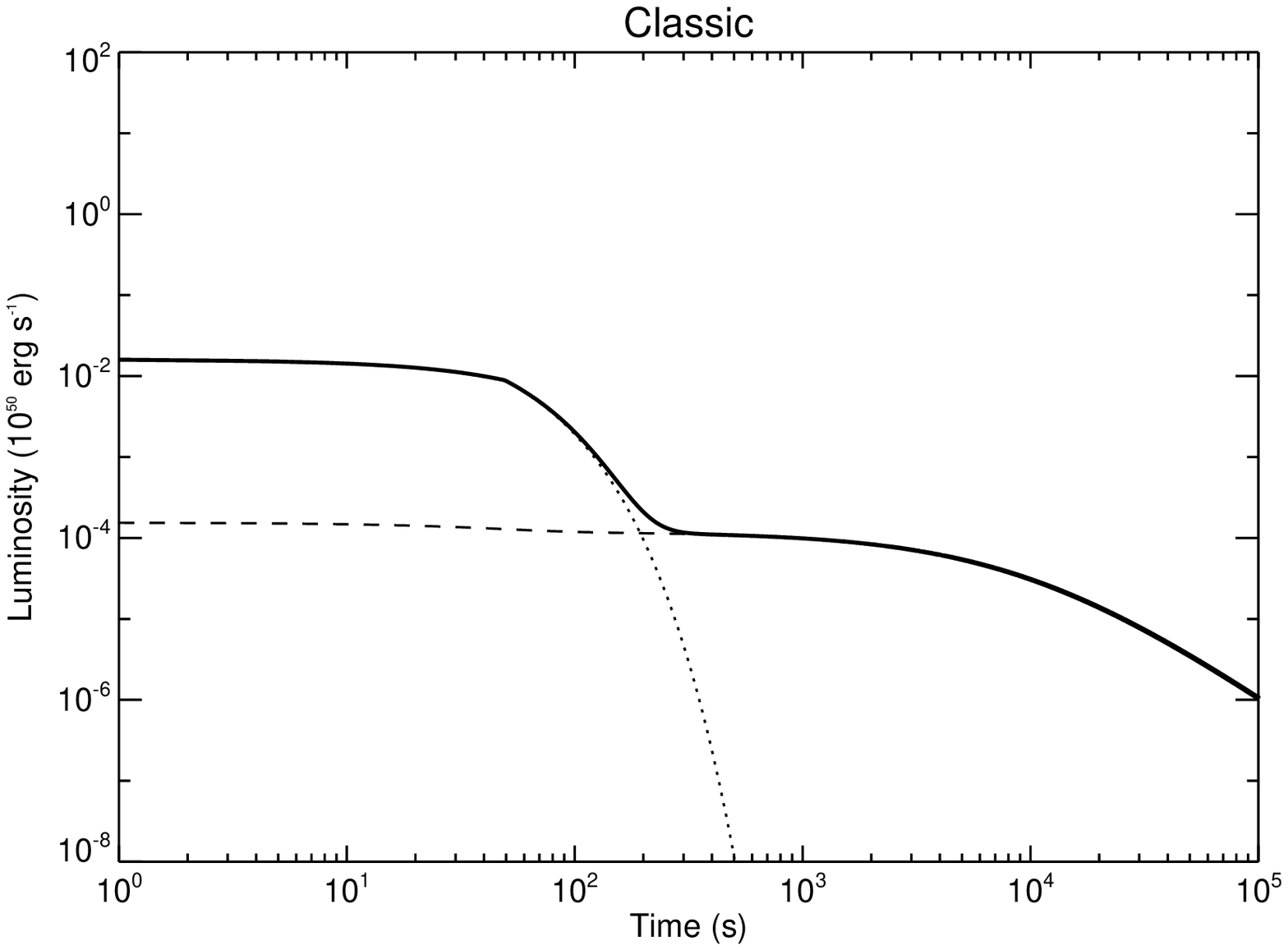}
\includegraphics[width=7.5cm,angle=-0]{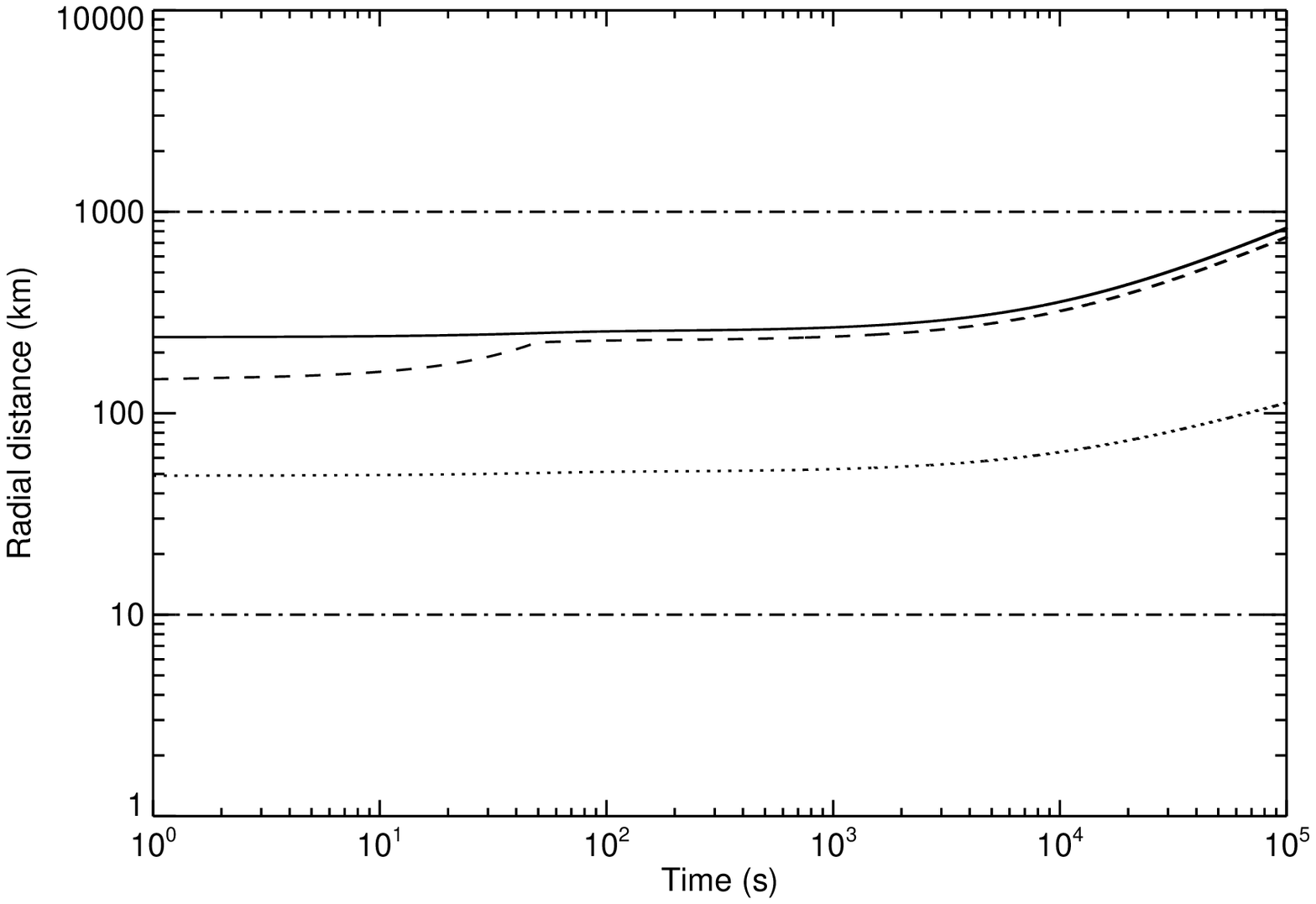}
\includegraphics[width=7.5cm,angle=-0]{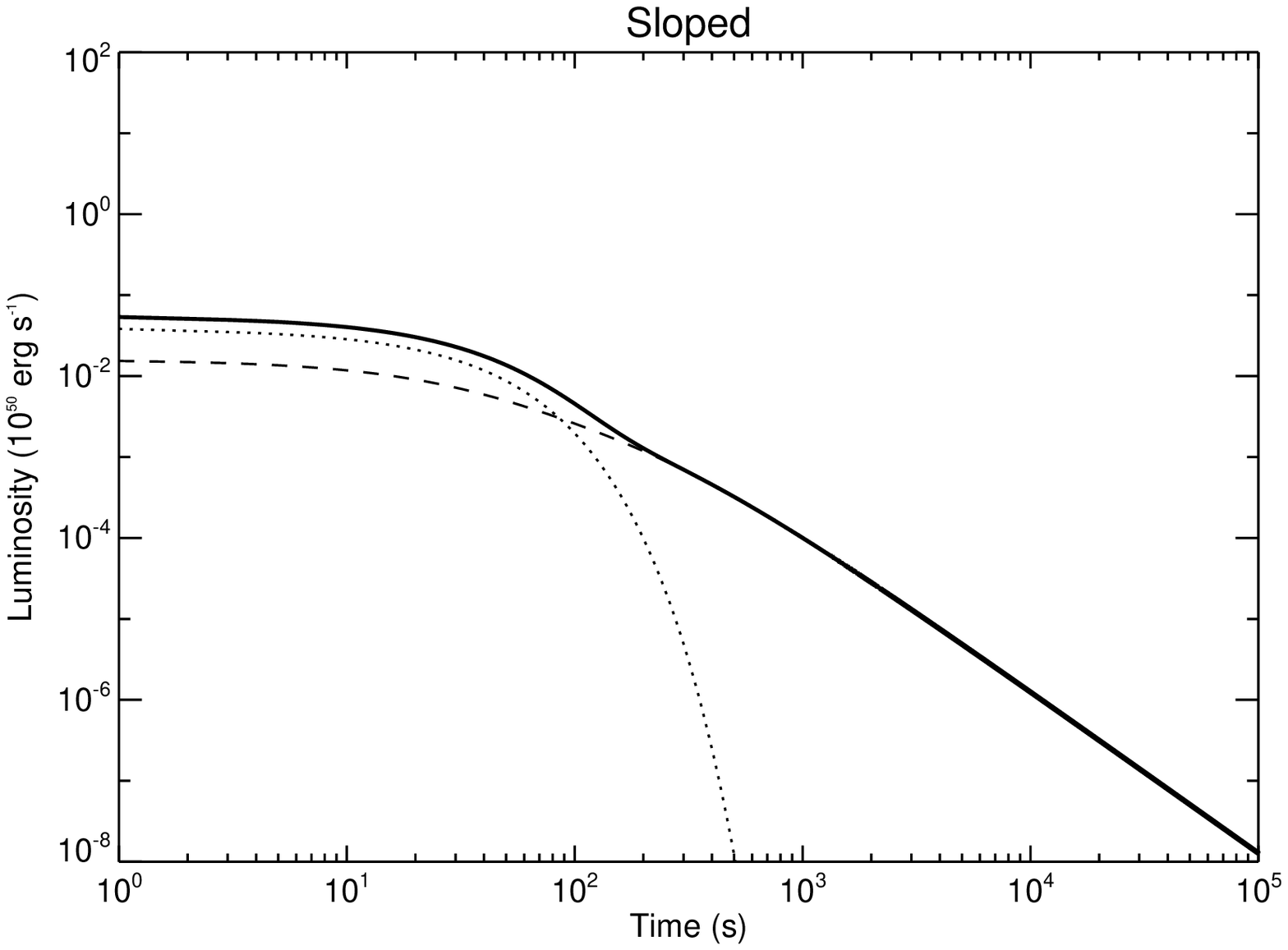}
\includegraphics[width=7.5cm,angle=-0]{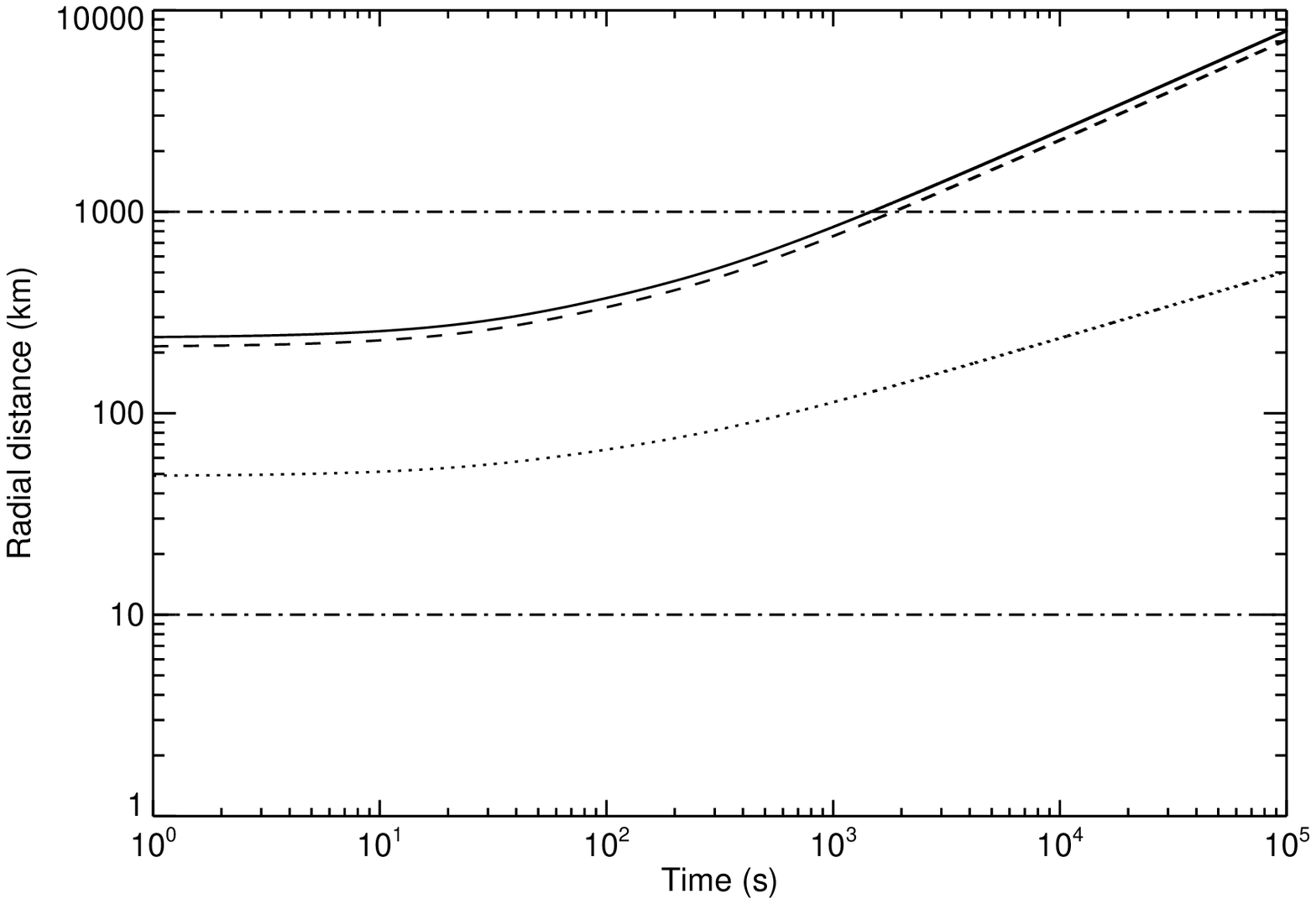}
\includegraphics[width=7.5cm,angle=-0]{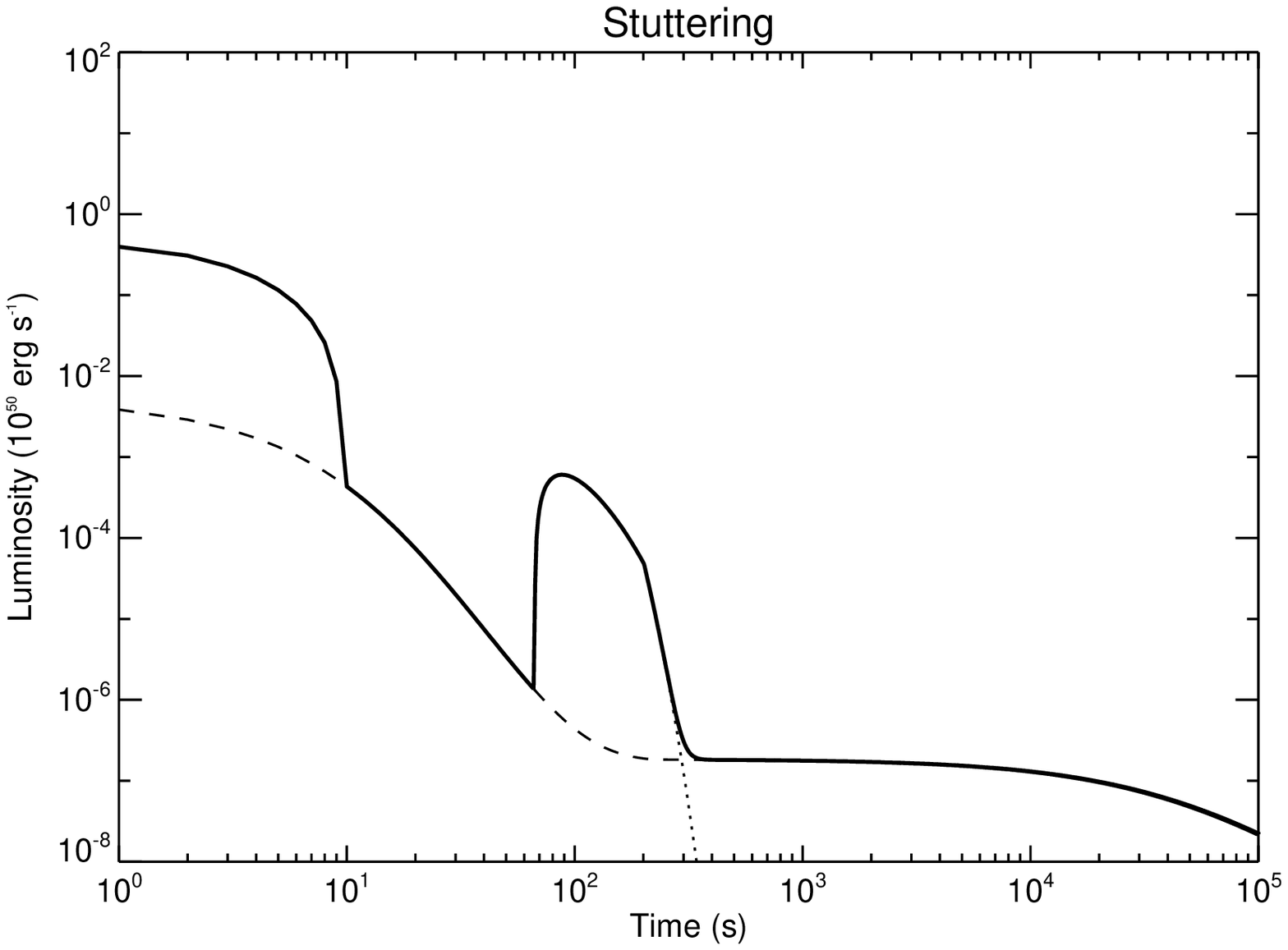}
\includegraphics[width=7.5cm,angle=-0]{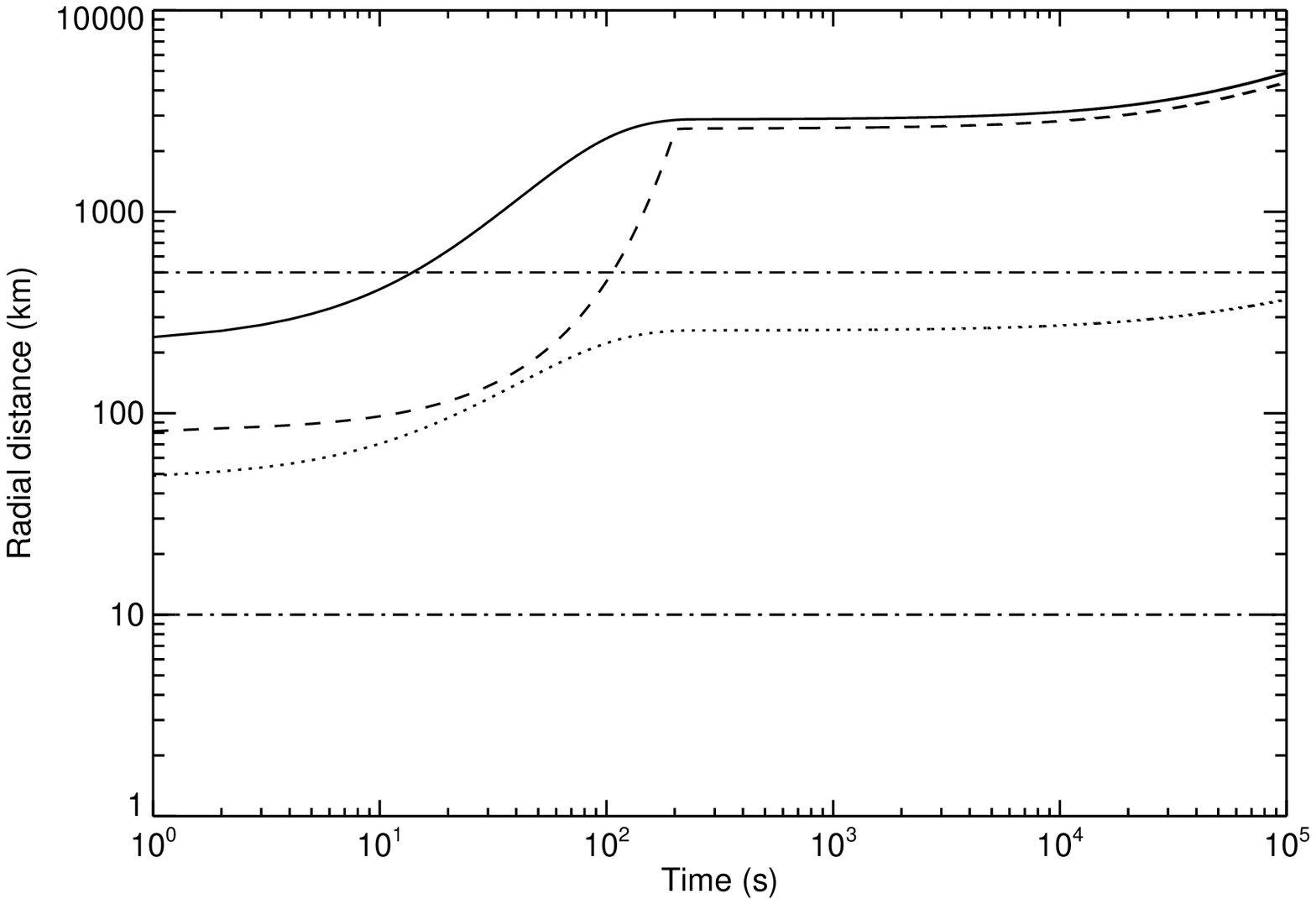}
\end{center}
\caption{Top to bottom: Type I - `Humped'; Type II - `Classic'; Type III - `Sloped'; Type IV - `Stuttering'. Each row shows plots for one example of one class. Intended to highlight phenomology only; they are not fully representative of the full range of morphology or energetics of their respective classes. These light curves do not contain the prompt spike. Left: Synthetic light curves representing the four identified phenomological classes. Dotted line - propeller luminosity. Dashed line - dipole luminosity. Right: Dotted (dashed) line shows the position of the co-rotation (Alfv\'{e}n) radius in km against time. Solid line shows the position of the light cylinder radius. Lower dot-dash line is the magnetar radius, upper dot-dash line is the outer disc radius.}
\label{templates}
\end{figure*}

\section{Testing parameter space}
To fully understand the morphological effects of the various parameters in the propeller model, the values given in Table 1 were assembled in every possible combination of spin period, $P$, magnetic field, $B$, disc mass $M_d$, disc radius, $R_d$, disc viscosity, $\alpha$, sound speed in the disc $c_s$ and NS mass, $M_*$. The result was a sample of 8100 synthetic light curves. We assume a constant $B$ throughout the duration of each, and the efficiency of both the propeller and dipole emission was set at 100\% since it serves only to normalise the luminosity in these cases. $k$ was set to $0.9$.

Initially, 540 light curves (all combinations of $P$, $B$, $M_d$, $R_d$ and $M_*$ with a constant $\alpha = 0.1$ and $c_s = 10^7$ cm s$^{-1}$) were examined in order to determine a classification system based on phenomology. After inspection, 4 clear types were identified. Example light curves of each type can be seen in Figure~\ref{templates}. Note that these light curves do not contain the prompt emission spike.

\subsection{Type I - `Humped'}
A `Humped' burst is born without propellering, initially powered by dipole emission alone (although this would be hidden beneath the prompt emission). As they progress, conditions for the initiation of propellering are met, and the light curve is given a `hump' by the rapid rise to prominence of the propeller luminosity. Propellering can be delayed like this for one of two reasons:\\
\\
a) $\dot{M}$ is high and/or $P$ is low so that $r_c > r_m$ and the system is in the accretion regime.\\
\\
b) $\dot{M}$ is high and/or $P$ is low enough that material cannot escape the potential well and Equation 11 is negative, despite $r_m > r_c$.\\
\\
These two possibilities can be distinguished by their light curves; bursts with strong initial accretion display a rising dipole luminosity at early times as the magnetar is spun up, whereas bursts with propellers too weak to enable matter to escape the potential well have flat dipole luminosity profiles at early times (e.g. Figure~\ref{templates}). 152 of the 540 synthetic bursts (28\%) are type I.

\subsection{Type II - `Classic'}
The `Classic' type can be formed by some combination of almost all parameters. They exhibit a relatively flat and well-defined propeller plateau, transitioning into a relatively flat and well-defined dipole plateau. In the extremes of parameter space (eg very high $B$ and low $P$), the other types are usually more prevelant, but a type II can still be formed given the right conditions. The division between this class and the type I or III bursts is rather loose, highlighting the smooth transition of parameters into `extreme' regimes. This class could also be further sub-divided into those experiencing rapid spin-down (shown by descending $L_{dip}$ at early times) and those which are comparatively stable (flat $L_{dip}$ at early times, see Figure~\ref{templates}). The divide between these is a combination of initial spin $P$ and the properties of the accretion disc; fast spinners spin down more rapidly, particularly when $\dot{M}$ is high, as this boosts the accretion torque. 202 of the 540 synthetic bursts (37\%) are type II.

\subsection{Type III - `Sloped'}
`Sloped' bursts are the result of the dipole component contributing strongly or even dominating the light curve during the propeller regime. In these cases, the two emission components appear to act as one, resulting in a poorly defined dipole plateau and a single component look to the light curve. This comes about when $B$ is high and/or spin is rapid, which are the conditions required for strong dipole emission. These types actually have the most powerful potential propellers, which is shown when the disc is small or loosely bound; In these conditions, $L_{prop}$ can rise above the already highly luminous $L_{dip}$, creating the brightest type II (Classic) bursts seen. A sloped burst may not be recognised as extended when observed, and would instead be classified as either a LGRB or SGRB. If accretion discs with increasingly low masses are considered, this could be the dividing point between EE GRBs and SGRBs. 63 of the 540 synthetic bursts (12\%) are type III.

\subsection{Type IV - `Stuttering'}
Light curves in the final burst category begin with propellering like a type II, but this rapidly vanishes after a few tens of seconds. After a short dipole-only phase, again lasting a few tens of seconds, the propeller is reborn, creating a hump much like a type I. The main factors governing this behaviour are $B$ and $M_d$. A high disc mass means that $\dot{M}$ is initially high. Propellering can still occur, due to the high magnetic field, but spin is lost rapidly through the accretion torque until it is too slow to power effective propellering. At this point, $L_{prop}$ shuts off and the light curve proceeds on $L_{dip}$ alone. In the absence of propellering, the rate of spin-down is greatly reduced, so that as the accretion rate begins to drop, the propeller makes a revival in much the same way as the type I bursts do. If the prompt emission is particularly strong or lasts a long time, a type IV may be observationally indistinguishable from a type I. Of the 540 synthetic bursts, 68 are type IV (13\%).\\
\\
In addition to the four classes, a total of 22 bursts (4\%) did not produce detectable propeller emission (i.e. the emission was less luminous than that of the dipole), and a further 33 (6\%) were unclassified due to incoherent and unrealistic light curves. These were exclusively bursts with the maximum ($5 \times 10^{16}$ G) magnetic field, indicating that magnetic fields much greater than this probably do not create EE GRBs; even at this $B$, particular conditions are required to produce a light curve in the correct energy region.

The parameters $\alpha$ and $c_s$ were then re-introduced as variables. As expected, no new phenomological classes were identified, and no existing classes dropped out. The overall effect was a greater range of morphologies within each class, specifically a general shortening/contraction of propeller regimes with increased $\alpha$ and/or $c_s$, and a slight elevation of peak luminosity.

It is clear from the results in Figure~\ref{templates} that the propeller model is capable of producing a variety of phenomena similar to those seen in GRB light curves, and given that we have restricted the behaviour of the fallback disc by requiring that it is fully formed and accreting at $t=0$, it seems likely the range is even greater. \citet{Piro11} have investigated the role propellering might play in the supernovae that power LGRBs, and \citet{Bernardini13} suggest it as a source of the precursor emission seen in some of the BAT6 sample. In addition, the smooth nature of propeller emission means it could conceivably reproduce the giant flares seen in some bursts (e.g. \citealt{Burrows05}). It may also be capable of uniting SGRBs with EE GRBs, as discussed in section 5.

\section{Fitting to observation}

The data sample to be used in fitting was taken from \citet{Gompertz13}. We include only bursts for which a value for $B$ and $P$ were found, and again assume a constant $B$ for the duration of each light curve. Table 2 lists the sample of 9 EE GRBs used. The model was written in {\sc idl} and made use of {\sc mpfit} \citep{Markwardt09}. Initial guesses for $B$ and $P$ during fitting were taken from \citet{Gompertz13}\footnote{We use corrected values for $B$ and $P$; an error was discovered in the k-correction calculations that means the results in that paper work out too high by a factor of (1+z) in Luminosity.}. These parameters were left fixed, leaving a 2 parameter fit comprised of $M_d$ and $R_d$. If no suitable fit was obtained then $P$ was set as a free parameter. If a fit was still not forthcoming, $B$ was unfrozen and allowed to vary as well. For all fits, The central magnetar was $1.4$ $M_{\odot}$ with a radius of $10$ km. $\alpha$ was held at $0.1$ and $c_s$ as $10^7$ cm s$^{-1}$. The conversion efficiency of kinetic energy to EM radiation in propellered material was set to 40\%, and the dipole efficiency to 5\%. $k$, the maximum fraction of $r_{lc}$ allowed for $r_m$, was $0.9$. Some flares were excluded from the fits. One at around 1000 seconds in GRB 070714B, and more noticably the late-time giant flare in GRB 050724.

The results of the fitting process can be seen in Table 3 and Figure~\ref{fits}. The light curves in Figure~\ref{fits} are a smoothed version of the original fit; once a fit was found, a plot was created using the resulting parameters running from $1$ to $10^5$ $s$ to show the global trend. In this way, the predicted behaviour from the fit can be observed during gaps in the light curve data. All light curves and associated results represent the global minimum $\chi^2$ value.

Figure~\ref{effplot} shows the effect that varying the efficiency of the propeller has on the result for $P$, the parameter most directly responsible for the luminosity output. It shows that for most bursts, efficiencies less than 10\% require a spin period more rapid than that of the break up frequency for the magnetar. This can be compensated for somewhat by varying $B$, $M_d$ and $R_d$ and exploring other regions for parameter space, but the general message is clear: the conversion of kinetic energy to EM waves must be fairly efficient for magnetic propellering to be succesful. Efficiencies of less than $\sim$ 10\% will not produce the required luminosity.

\begin{figure*}
\begin{center}
\includegraphics[width=7.7cm,angle=-0]{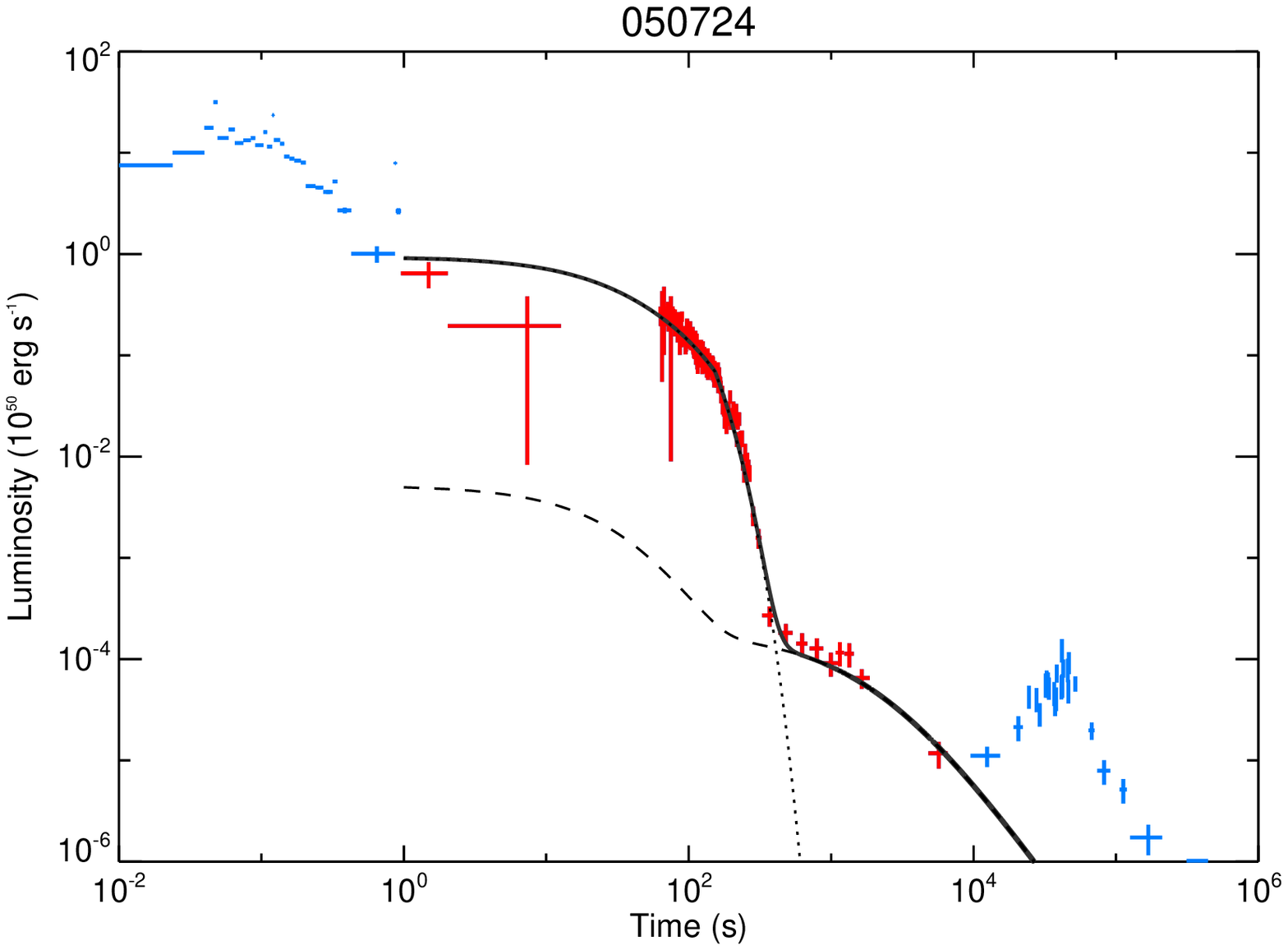}
\includegraphics[width=7.7cm,angle=-0]{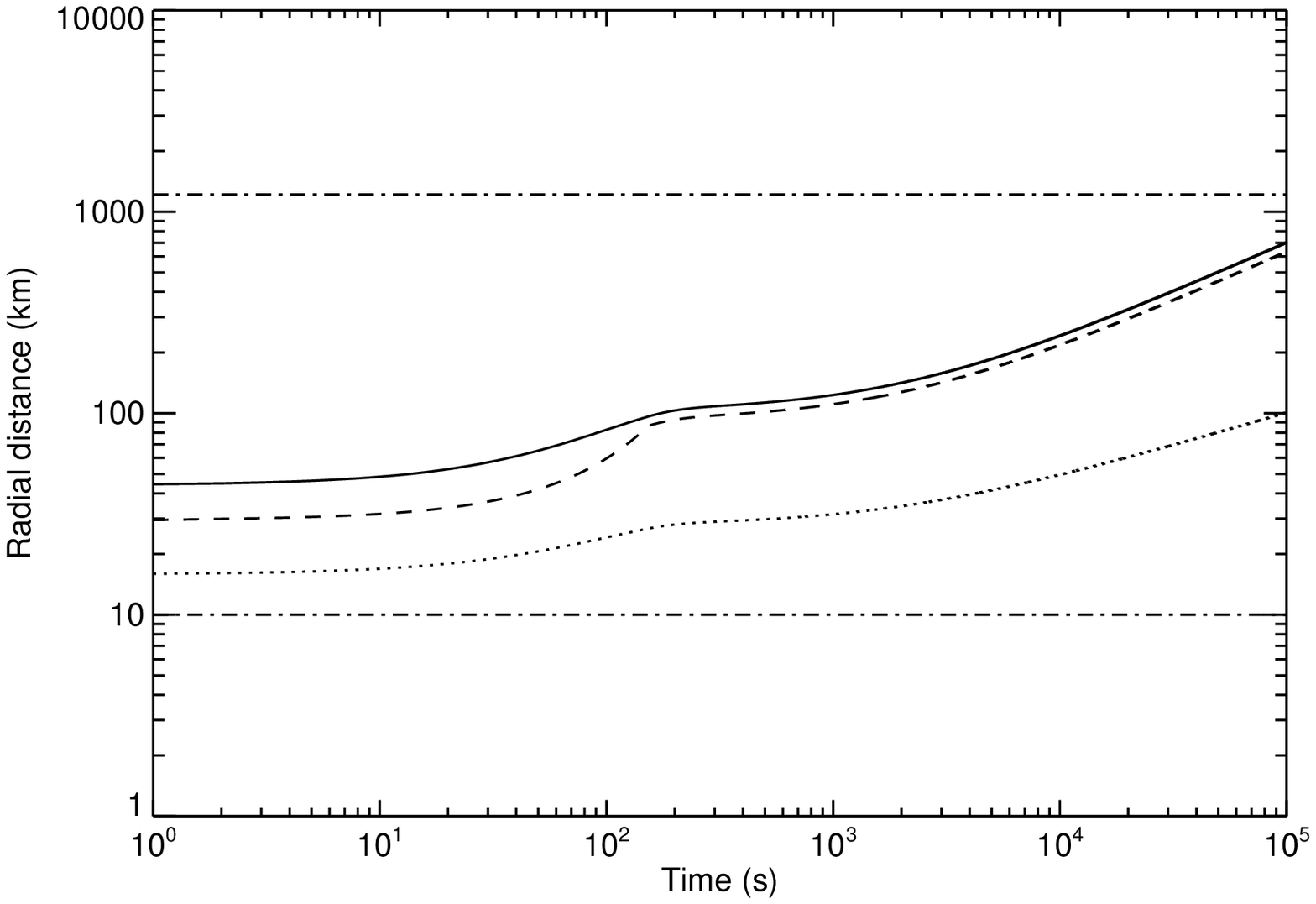}
\includegraphics[width=7.7cm,angle=-0]{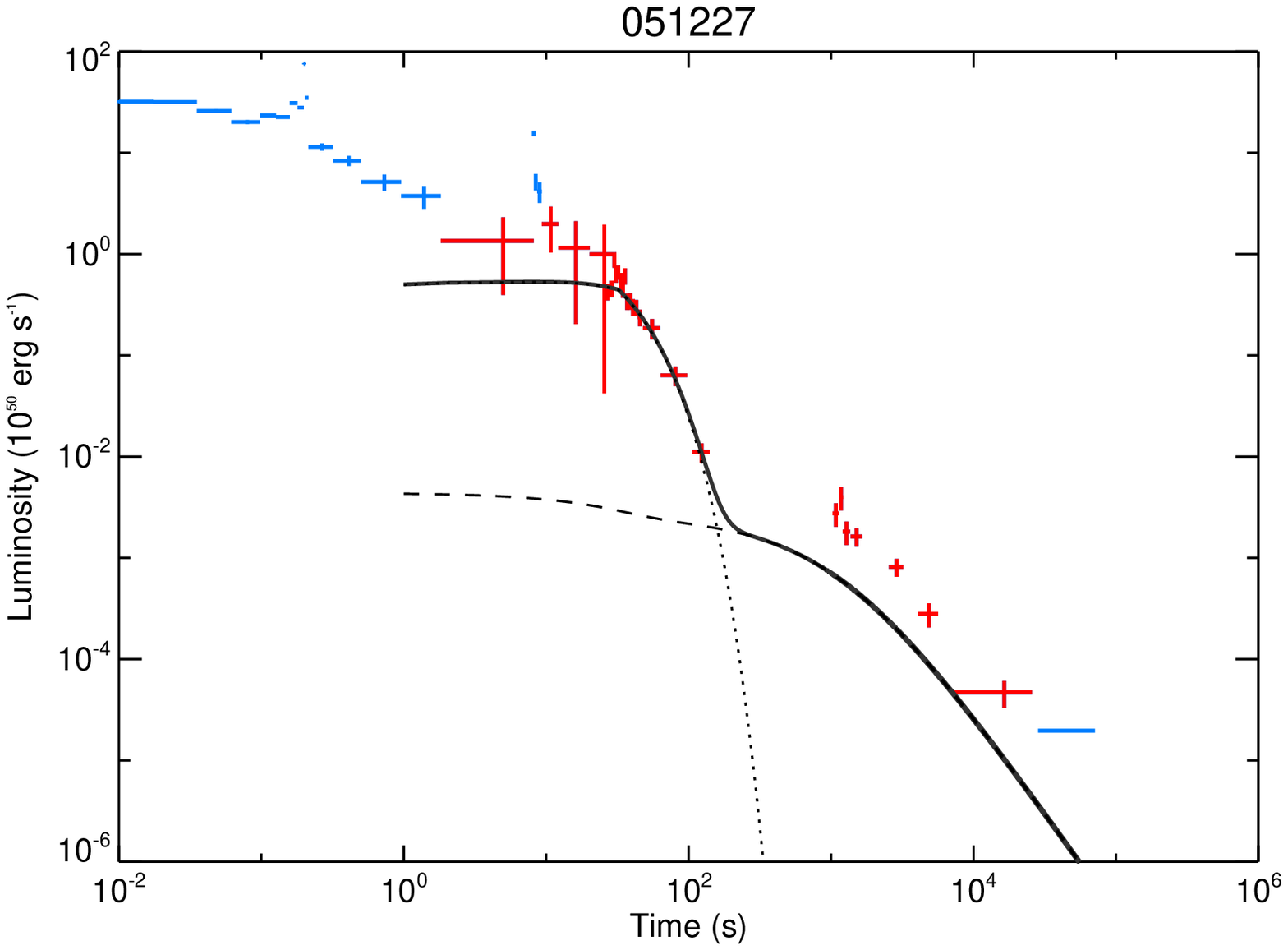}
\includegraphics[width=7.7cm,angle=-0]{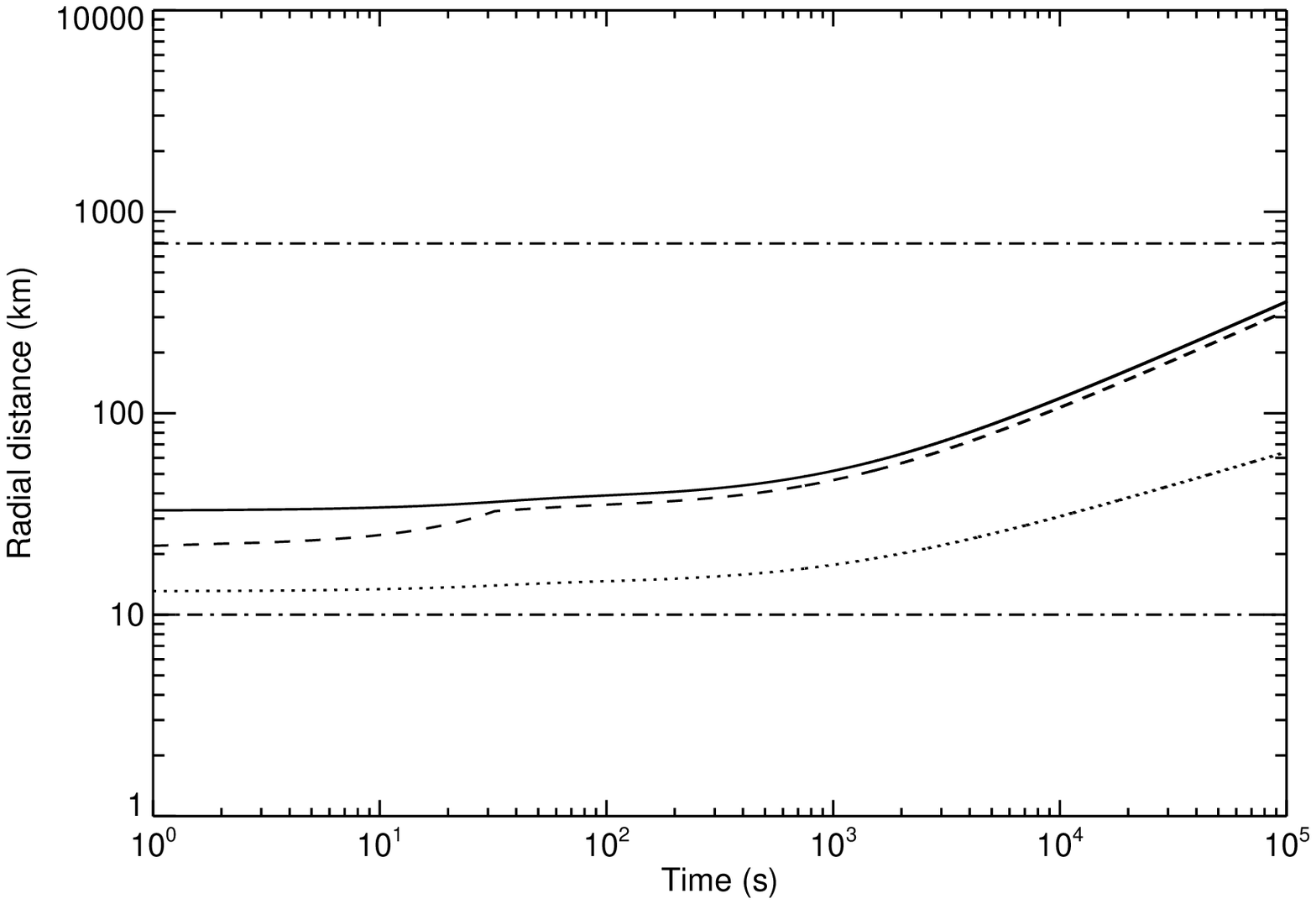}
\includegraphics[width=7.7cm,angle=-0]{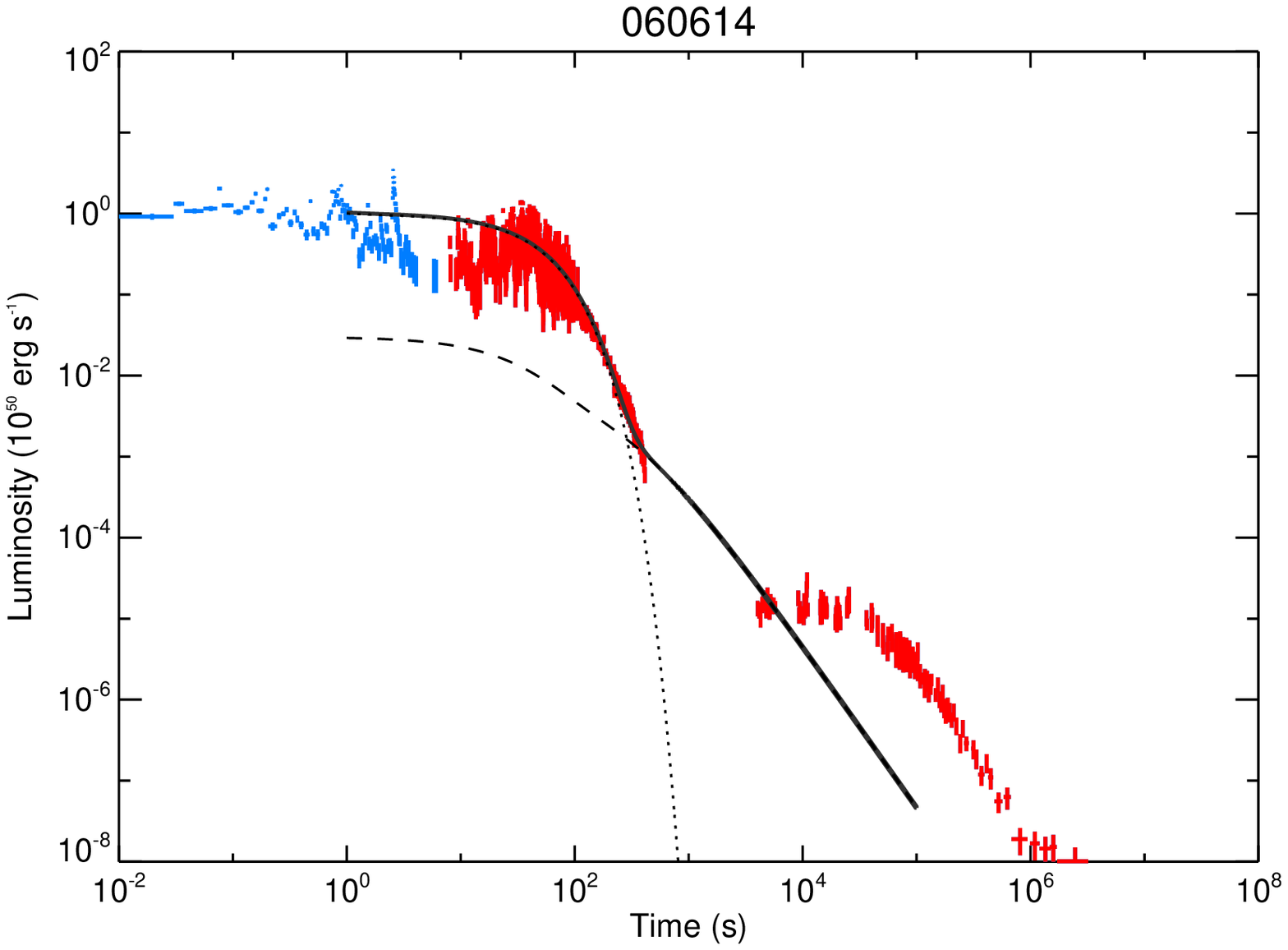}
\includegraphics[width=7.7cm,angle=-0]{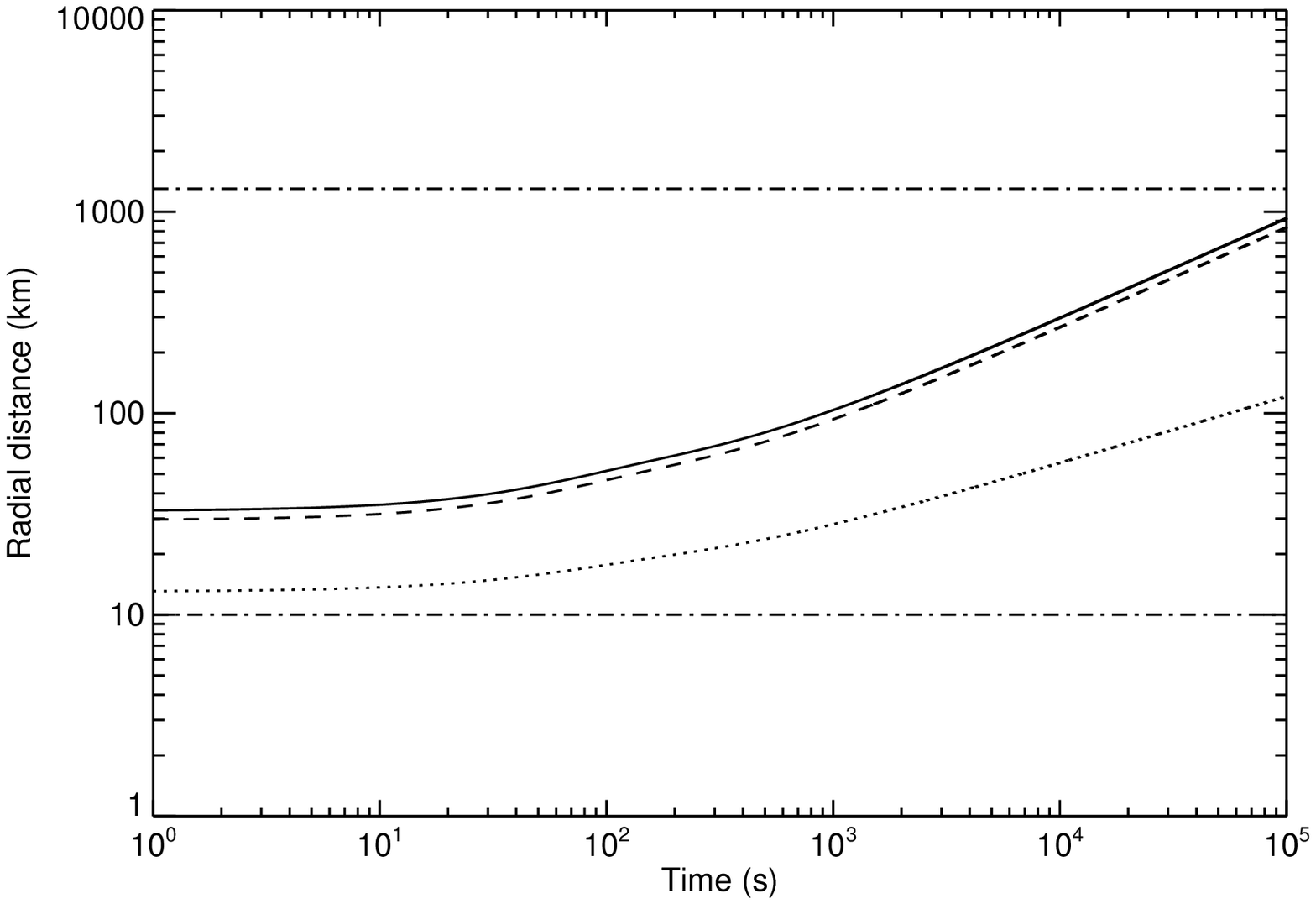}
\includegraphics[width=7.7cm,angle=-0]{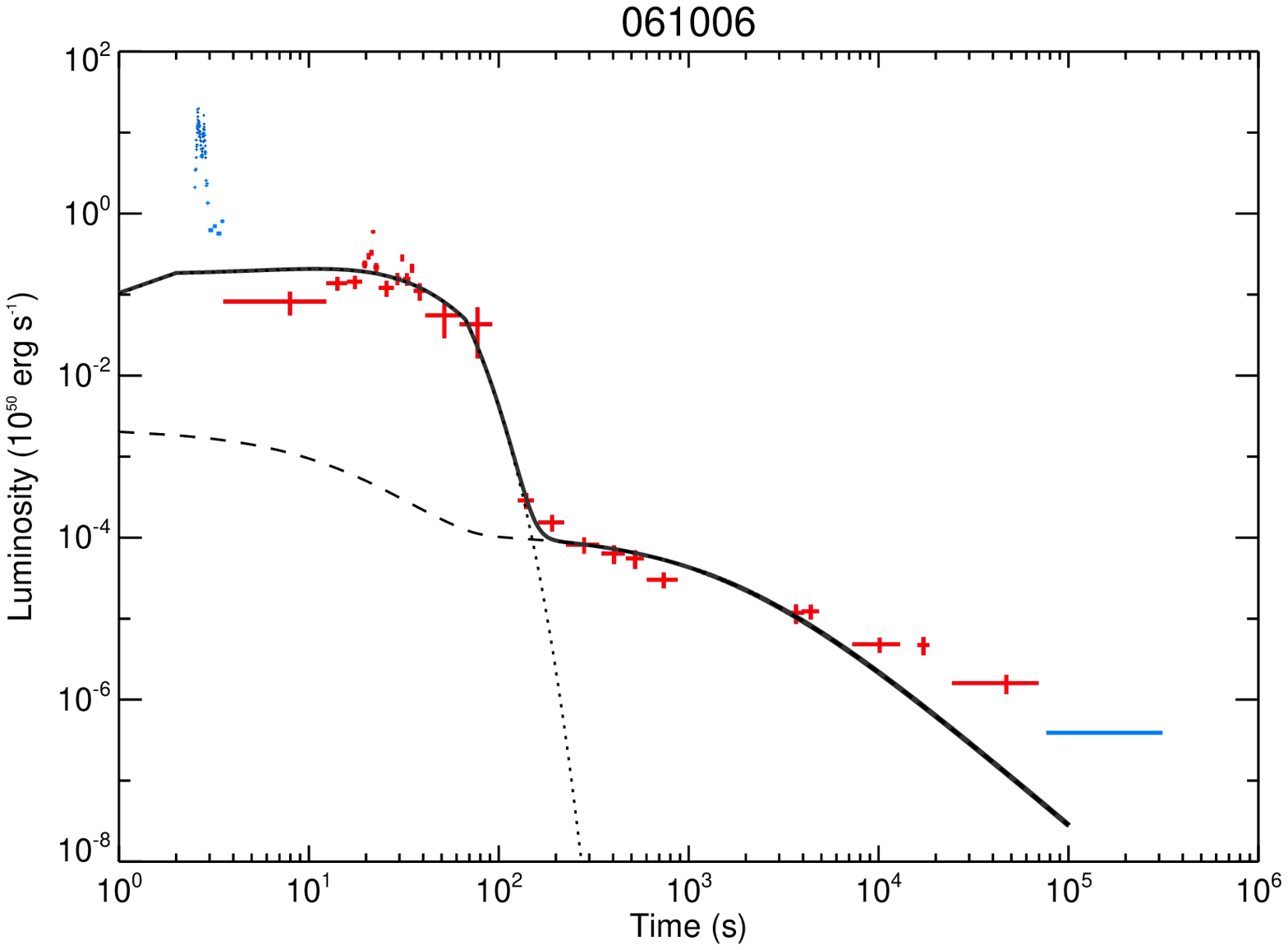}
\includegraphics[width=7.7cm,angle=-0]{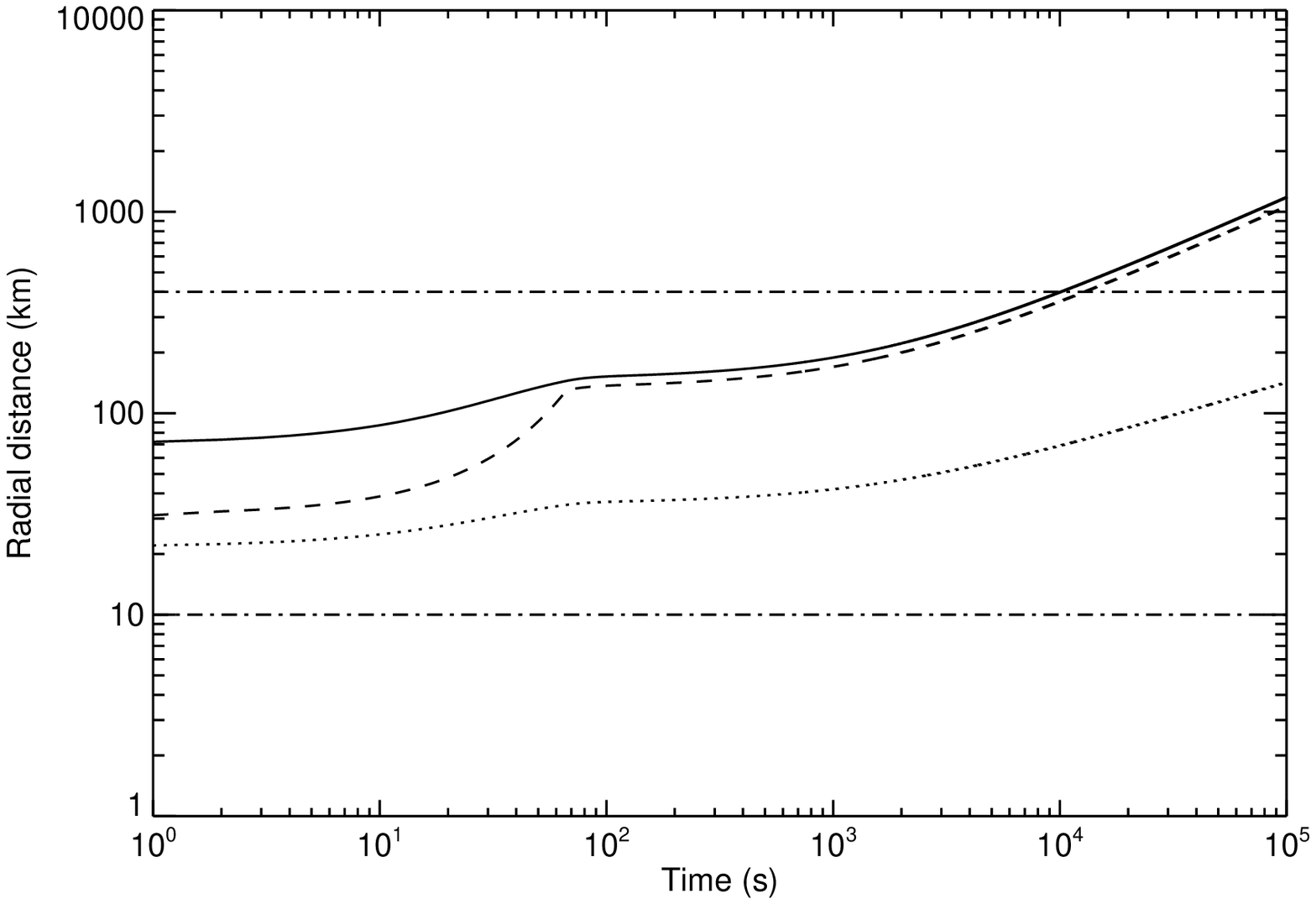}
\end{center}
\caption{Each row shows details for one burst. Left: Black line - model fit; Red points - data that has been fitted to; Blue points - data not fitted to. Right: Dotted (dashed) line shows the position of the co-rotation (Alfv\'{e}n) radius in km against time. Solid line marks the light cylinder radius. Lower dot-dash line is the magnetar radius, upper dot-dash line is the outer disc radius.}
\label{fits}
\end{figure*}
\addtocounter{figure}{-1}
\begin{figure*}
\begin{center}
\includegraphics[width=7.7cm,angle=-0]{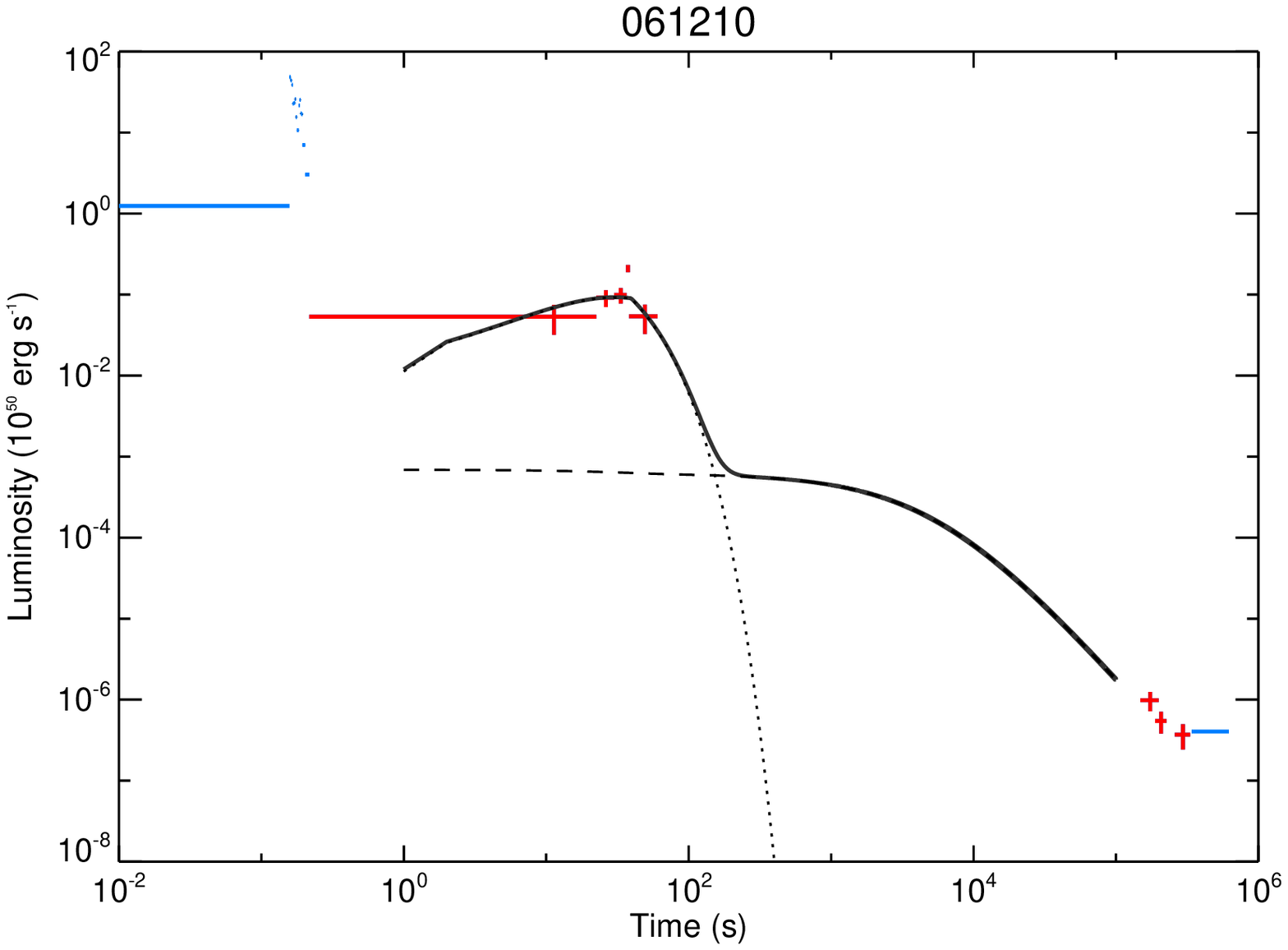}
\includegraphics[width=7.7cm,angle=-0]{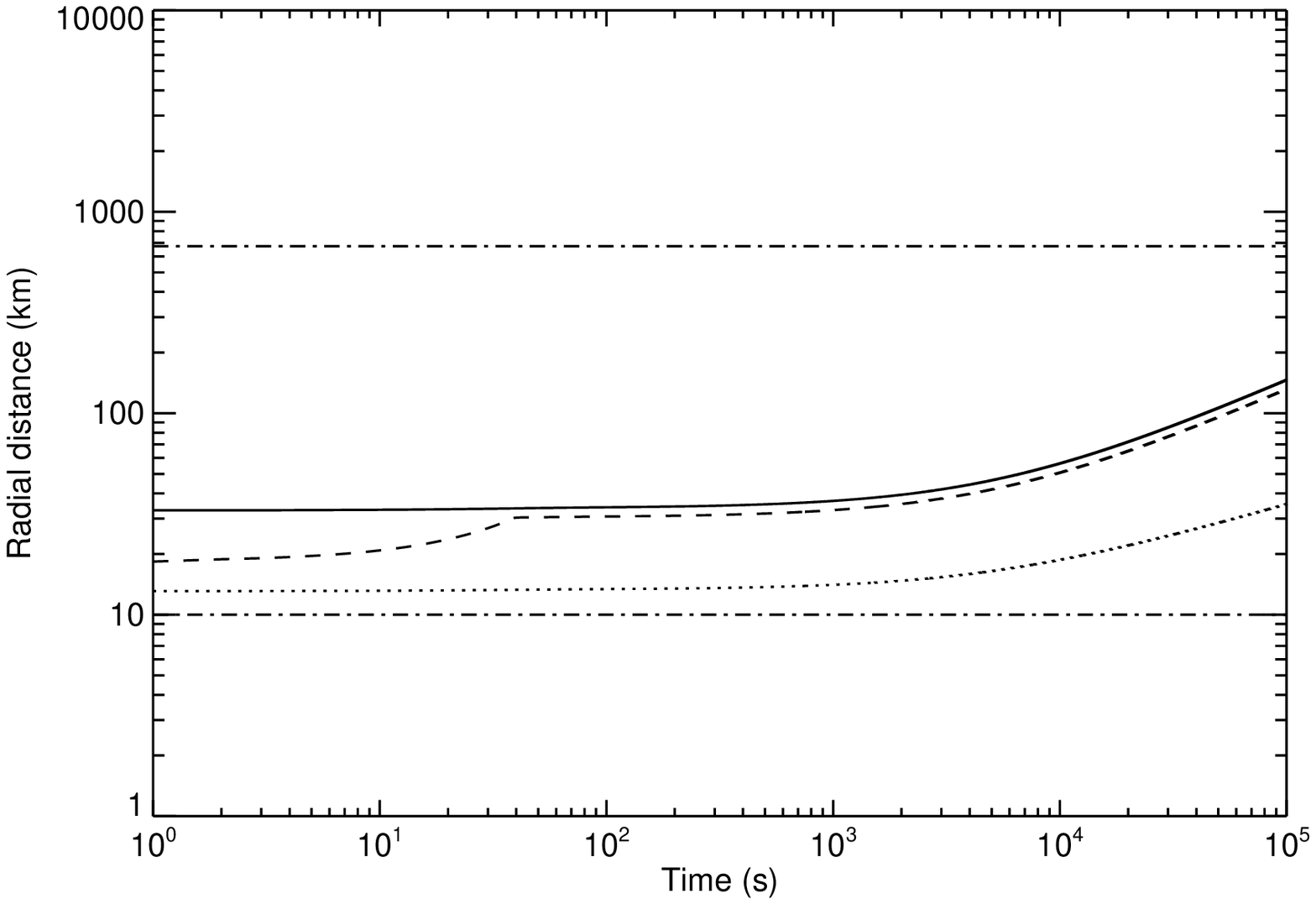}
\includegraphics[width=7.7cm,angle=-0]{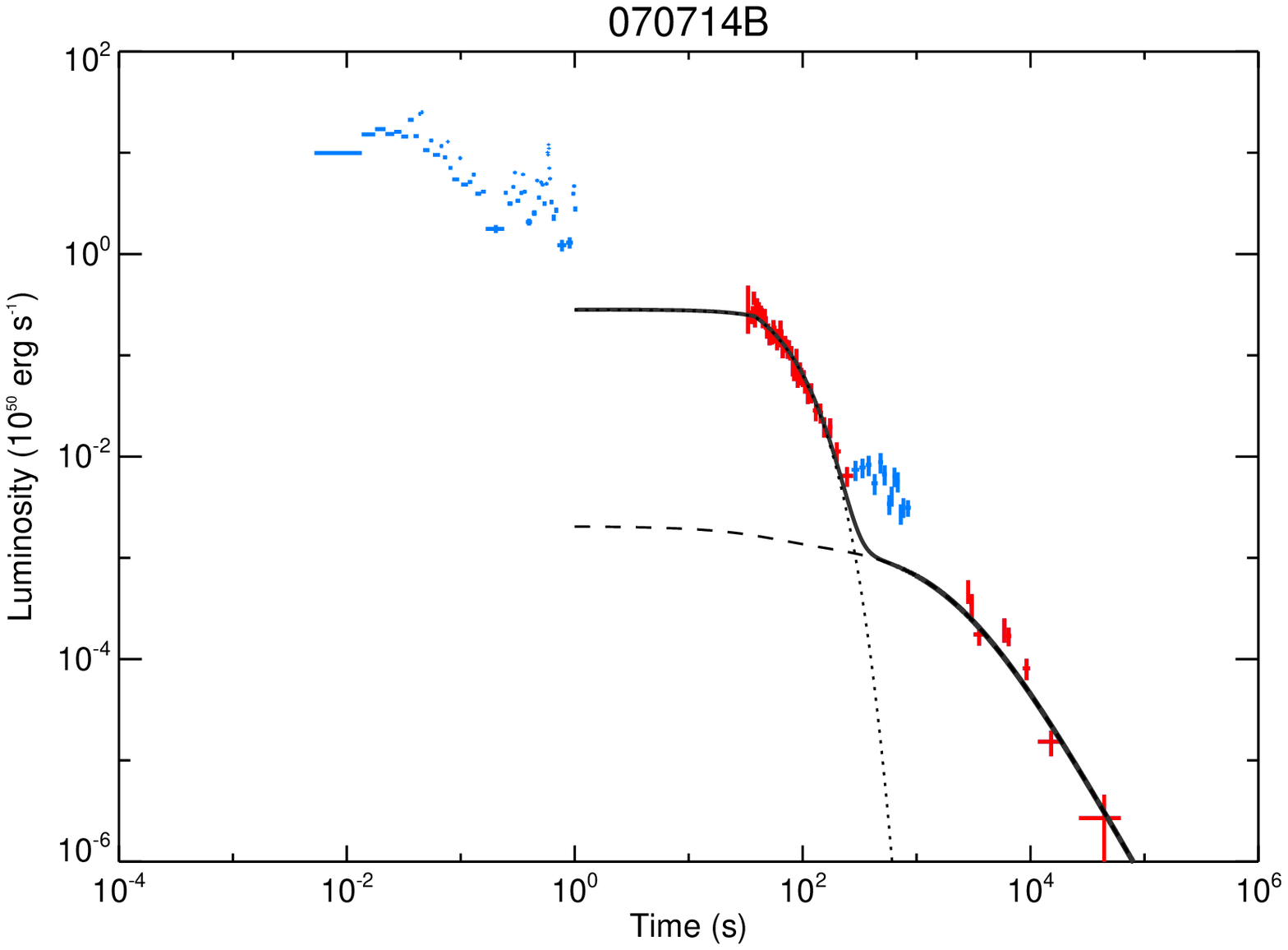}
\includegraphics[width=7.7cm,angle=-0]{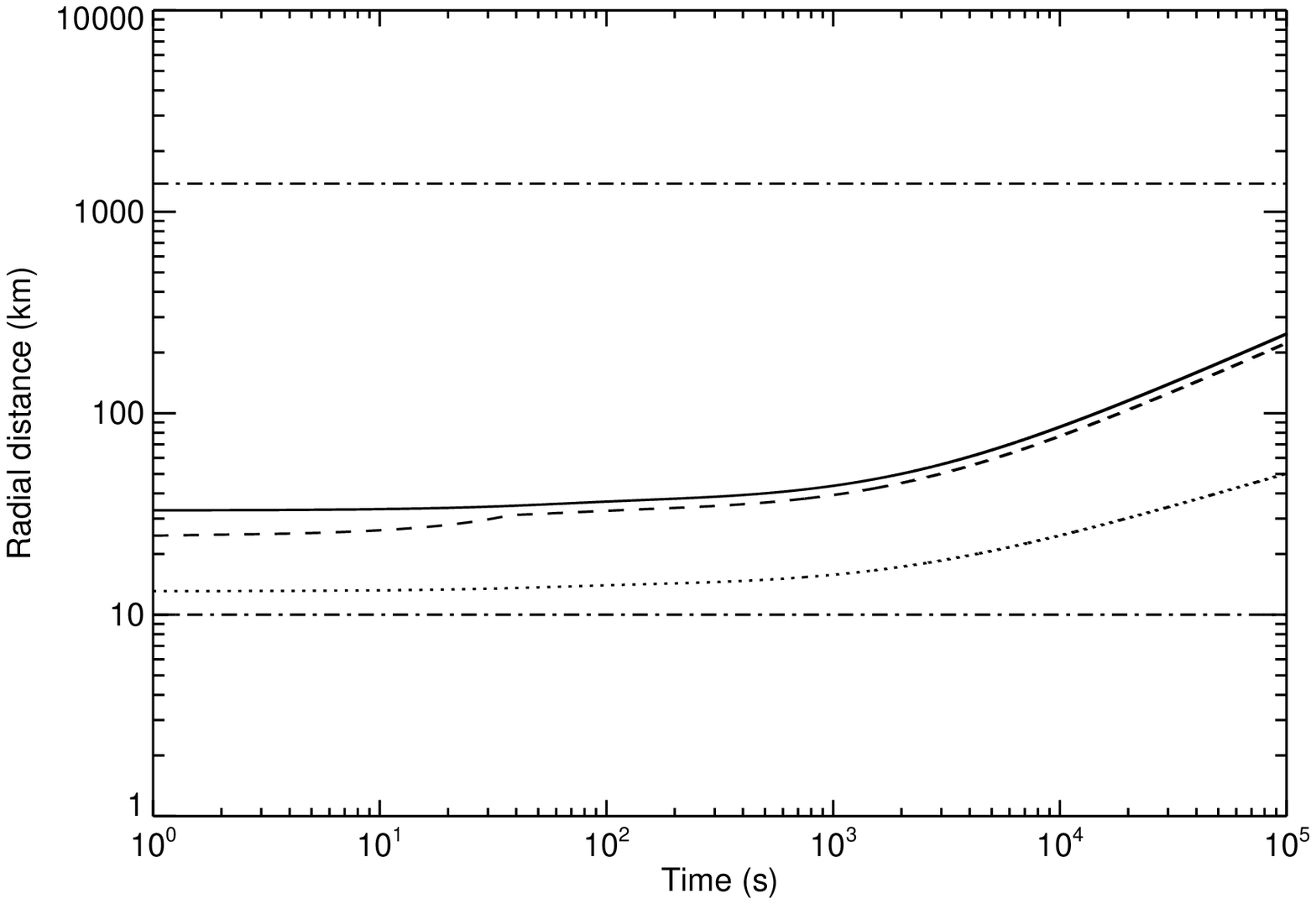}
\includegraphics[width=7.7cm,angle=-0]{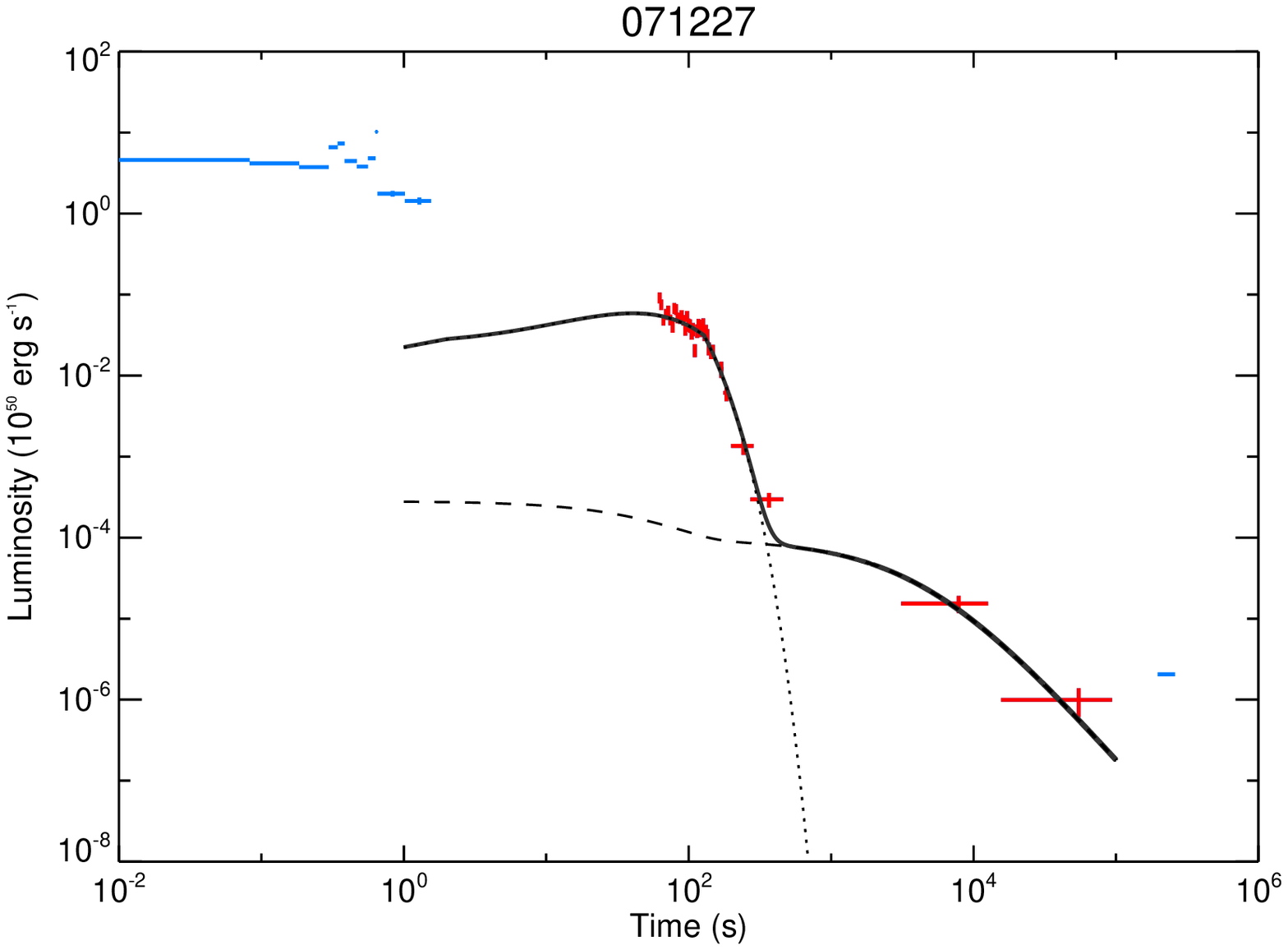}
\includegraphics[width=7.7cm,angle=-0]{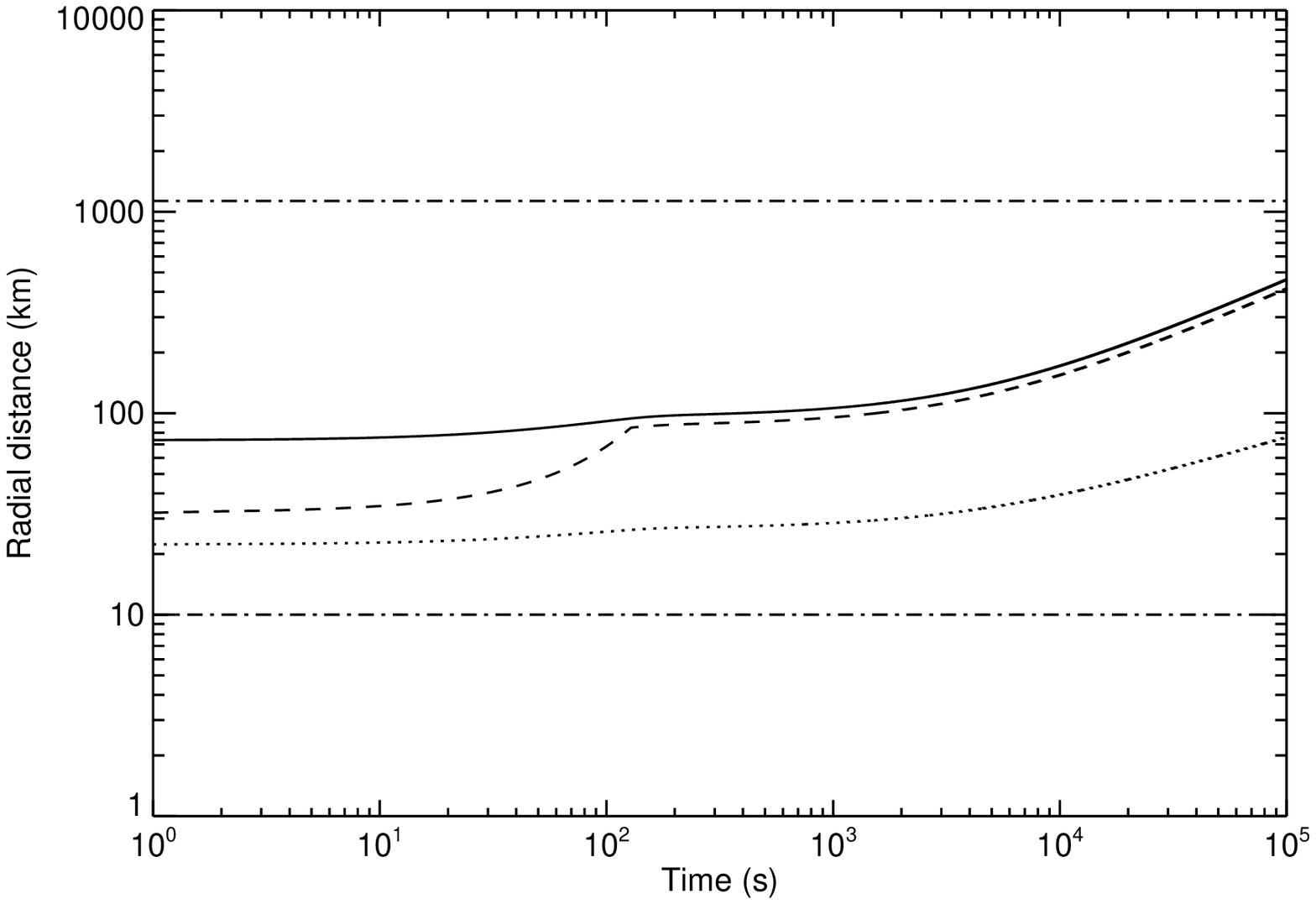}
\includegraphics[width=7.7cm,angle=-0]{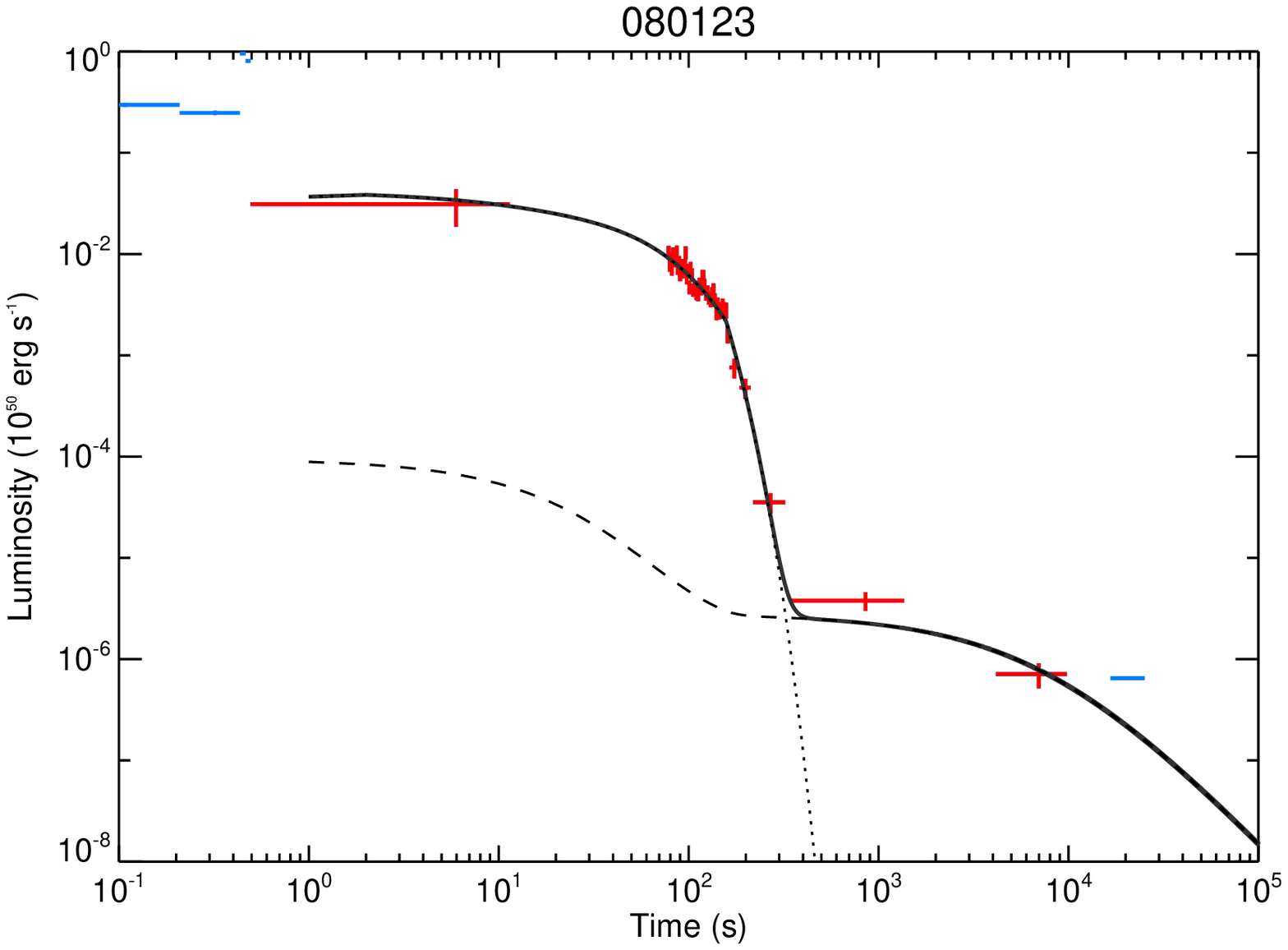}
\includegraphics[width=7.7cm,angle=-0]{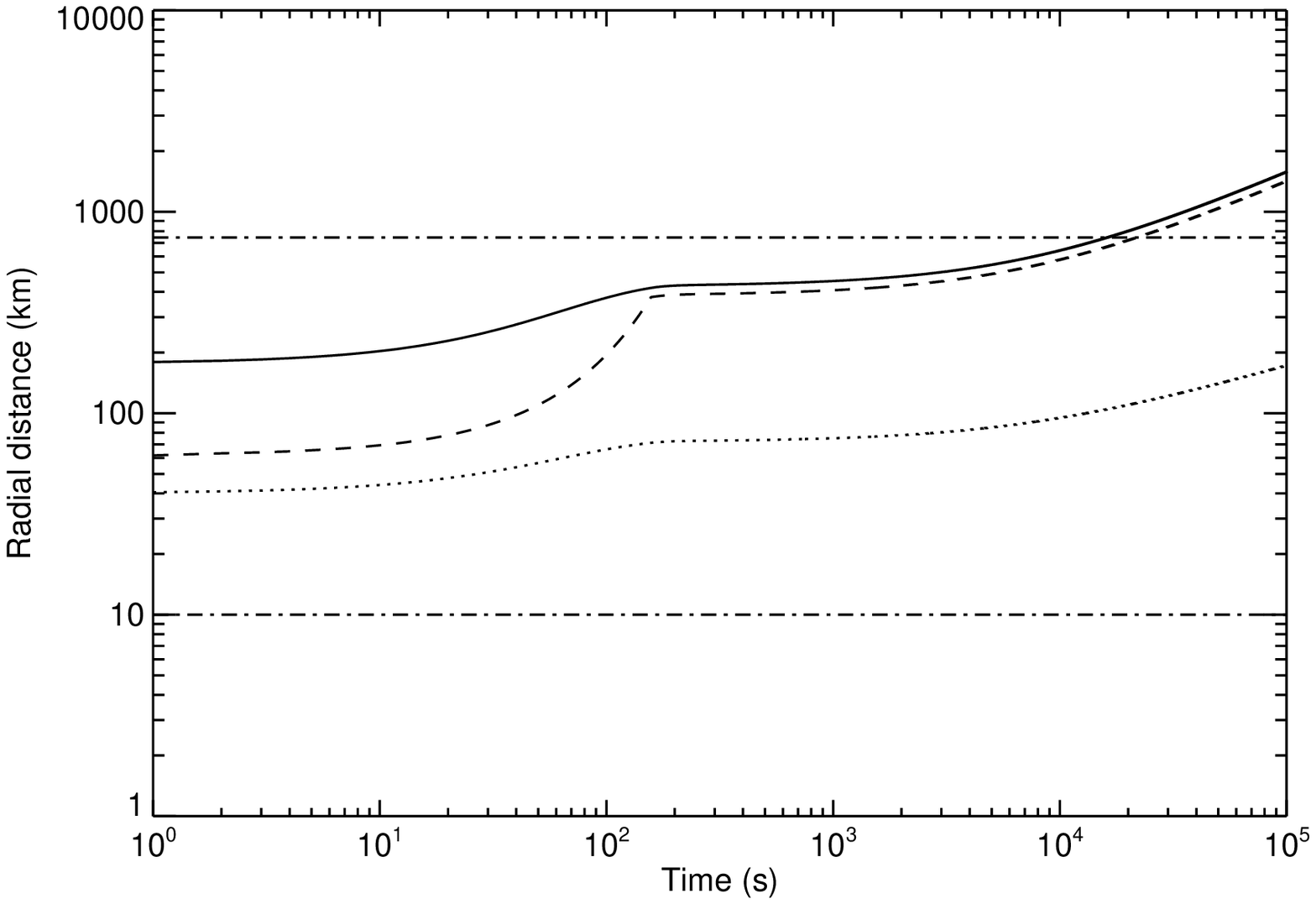}
\end{center}
\caption{\textbf{cont.} Each row shows details for one burst. Left: Black line - model fit; Red points - data that has been fitted to; Blue points - data not fitted to. Right: Dotted (dashed) line shows the position of the co-rotation (Alfv\'{e}n) radius in km against time. Solid line marks the light cylinder radius. Lower dot-dash line is the magnetar radius, upper dot-dash line is the outer disc radius.}
\end{figure*}
\addtocounter{figure}{-1}
\begin{figure*}
\begin{center}
\includegraphics[width=7.7cm,angle=-0]{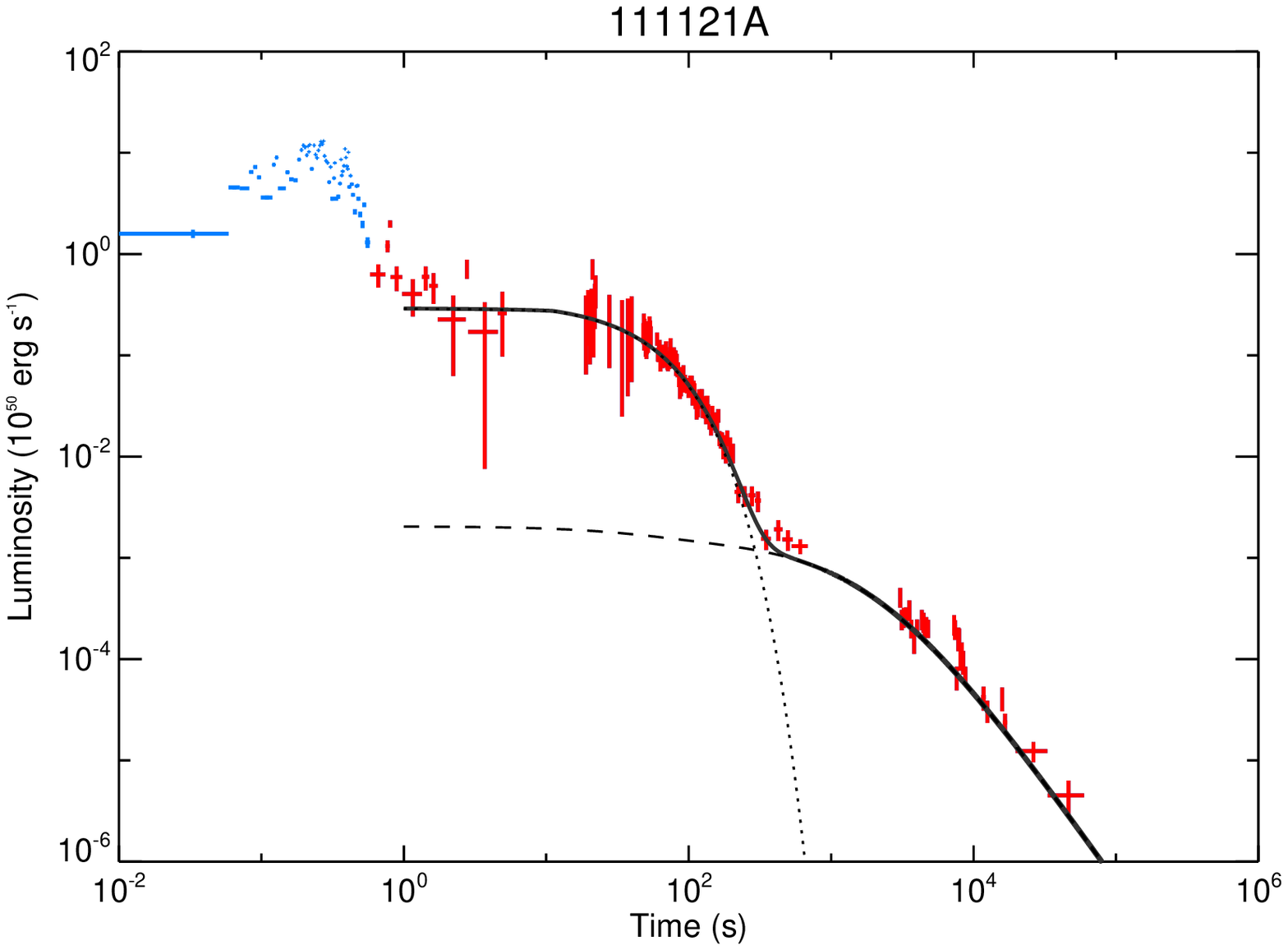}
\includegraphics[width=7.7cm,angle=-0]{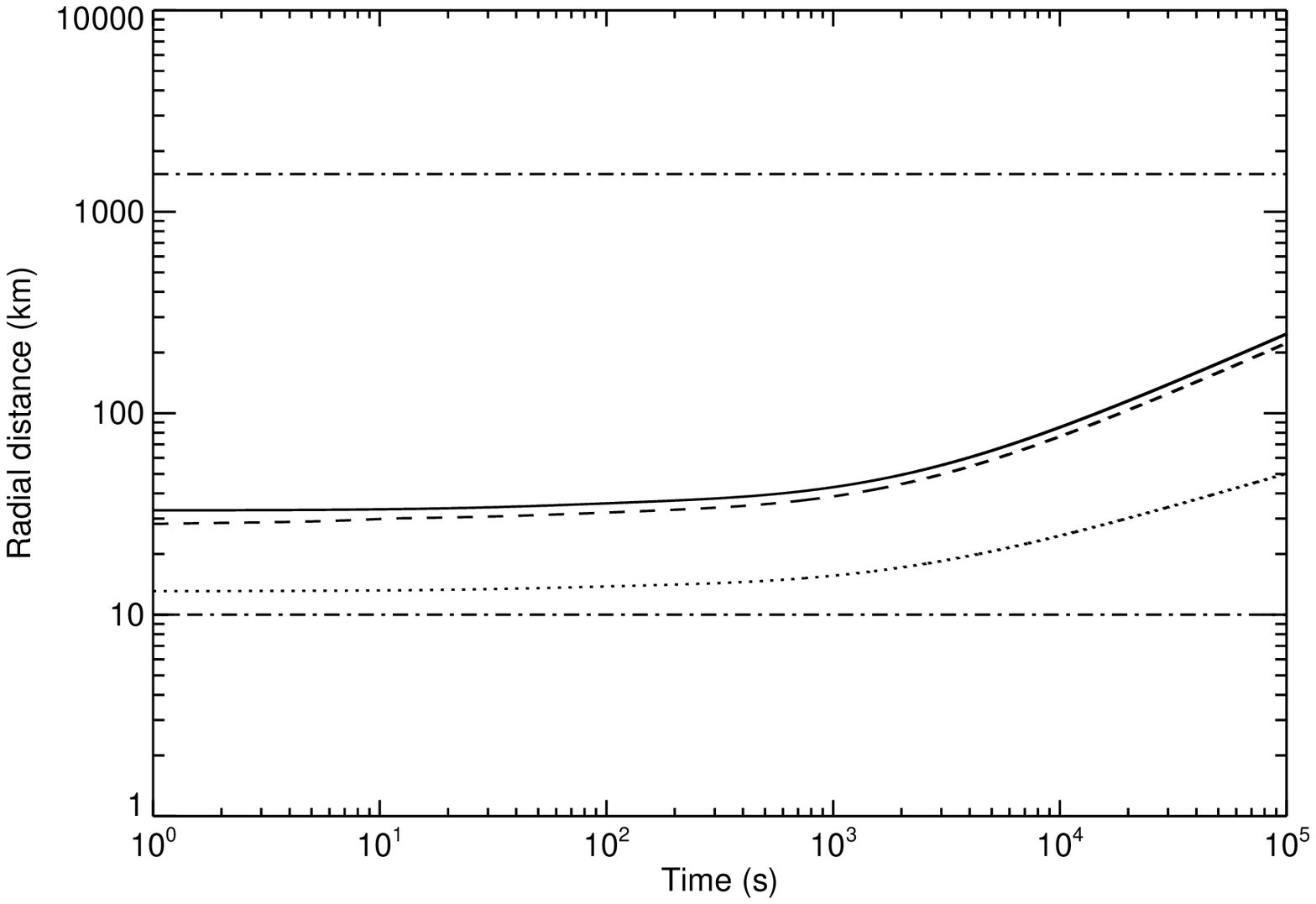}
\end{center}
\caption{\textbf{cont.} Each row shows details for one burst. Left: Black line - model fit; Red points - data that has been fitted to; Blue points - data not fitted to. Right: Dotted (dashed) line shows the position of the co-rotation (Alfv\'{e}n) radius in km against time. Solid line marks the light cylinder radius. Lower dot-dash line is the magnetar radius, upper dot-dash line is the outer disc radius.}
\end{figure*}

\begin{table}
\begin{center}
\begin{tabular}{lccl}
\hline
GRB & $\Gamma$ & z & Ref.\\
\hline
050724 & 1.77 & 0.2576$^1$ & \citet{Covino05} \\
051227 & 1.46 & 2.8$^{a,2}$ & \citet{Barbier05} \\
060614 & 1.79 & 0.1254$^3$ & \citet{Parsons06} \\
061006 & 2.03 & 0.4377$^4$ & \citet{Schady06} \\
061210 & 2.20 & 0.4095$^5$ & \citet{Cannizzo06} \\
070714B & 1.15 & 0.9224$^6$ & \citet{Racusin07} \\
071227 & 1.54 & 0.381$^7$ & \citet{Sakamoto07} \\
080123 & 1.99 & (0.39) & \citet{Ukwatta08} \\
111121A & 1.50 & (0.39) & \citet{D'Elia11} \\
\hline
\end{tabular}
\caption{Selected sample of EE GRBs. Bracketed values for redshift, $z$, indicate no published value was available. In these cases the mean value of the EE sample where $z$ is known was used. a - upper limit. 1 - \citet{Prochaska05}; 2 - \citet{D'Avanzo09}; 3 - \citet{Price06GCN}; 4 - \citet{Berger07}; 5 - \citet{Cenko06}; 6 - \citet{Graham09}; 7 - \citet{D'Avanzo07}}
\end{center}
\end{table}

\begin{table*}
\begin{center}
\begin{tabular}{lcccc}
\hline
GRB & P & B & $M_d$ & $R_d$ \\
 & (ms) & ($10^{15} G$) & ($M_{\odot}$) & (km) \\
\hline
050724 & 0.93 $\pm 0.04$ & 0.88 $\pm 0.04$ & (2.63 $\pm 0.13$) $\times 10^{-2}$ & 1217 $\pm 4$ \\
051227 & 0.69 [L] & 0.45 $\pm 0.19$ & (1.10 $\pm 0.18$) $\times 10^{-2}$ & 695 $\pm 41$ \\
060614 & 0.69 [L] & 1.17 $\pm 0.05$ & (1.20 $\pm 0.01$) $\times 10^{-2}$ & 1300 $\pm 4$ \\
061006 & 1.51 $\pm 0.21$ & 1.48 $\pm 0.07$ & (2.01 $\pm 0.37$) $\times 10^{-2}$ & 400 $\pm 2$ \\
061210 & 0.69 [L] & 0.18 $\pm 0.05$ & (3.20 $\pm 2.88$) $\times 10^{-3}$ & 674 $\pm 753$ \\
070714B & 0.69 [L] & 0.31 $\pm 0.05$ & (6.91 $\pm 0.28$) $\times 10^{-3}$ & 1378 $\pm 72$ \\
071227 & 1.54 $\pm 0.12$ & 0.57 $\pm 0.08$ & (7.63 $\pm 1.02$) $\times 10^{-3}$ & 1131 $\pm 17$ \\
080123 & 3.75 $\pm 0.46$ & 1.92 $\pm 0.16$ & (5.82 $\pm 1.10$) $\times 10^{-3}$ & 742 $\pm 6$ \\
111121A & 0.69 [L] & 0.31 $\pm 0.03$ & (4.80 $\pm 0.10$) $\times 10^{-3}$ & 1538 $\pm 43$ \\
\hline
\end{tabular}
\caption{Results from fitting the propeller model to 9 EE GRBs. Values with an [L] came up against the parameter limit for the minimum allowed spin period, and therefore do not have associated errors. Errors are $1\sigma$.}
\end{center}
\end{table*}

\section{Discussion}
Our derived accretion disc masses and radii are all broadly consistent with theoretical predictions \citep{Lee09}, lying in the range of a few $10^{-3}$ $M_{\odot}$ to a few $10^{-2}$ $M_{\odot}$ and $\sim$ 400 km -- 1500 km respectively. For only one burst, GRB 071227, is the initial spin period consistent with that in \citet{Gompertz13}. This is not surprising; the two studies were done with different efficiencies for the dipole (5\% in this study vs 100\% in \citealt{Gompertz13}), and the rate of spin-down was enhanced by the inclusion of an accretion disc (Equation 8) which was not present in the previous work.

Most bursts in the sample show evidence of a smooth connection from the prompt to EE phase, however GRB 060614 and GRB 061006 appear to struggle to capture the rising profile of propellering at early times. This could be explained by the simplicity in the model used; we had to assume the accretion disc was present at $t = 0$ seconds, meaning accretion began immediately and at its peak strength. In reality, material would still be falling back at this time, so that accretion would initially be much gentler, but would grow in strength as the disc was fed. \citet{Lee09} predict the material would return on a timescale of $\sim 10$ seconds, which would help explain these features.

The model has some trouble fitting the extended tail and dipole plateau in GRB 060614 simultaneously. The problem is caused by the longer than normal plateau, which turns over at around $10^5$ s rather than the $10^3$ -- $10^4$ s seen in the other bursts. Sustaining the plateau for this long requires a low value for $B$ ($\sim 10^{14}$ G) or a long spin period ($\sim 10$ ms), but the very luminous extended tail in this burst demands exactly the opposite. Fitting tends to favour the demands of the EE, since this is where more of the data points are found. The problem can be partially solved by varying the efficiency between the two components, since this has the effect of increasing or decreasing the power-law slope that connects them, but for reasonable values of efficiency, a discrepancy still remains. One potential solution to this dilemma is the possibility that the magnetic field is not constant in this burst (or, probably more accurately, more varying than the other bursts). An order of magnitude decay in the magnetic field can extend the duration of the dipole plateau by more than an order of magnitude temporally, more than enough for the requirements of 060614, although how, and if any emission would be observable is unclear. The model can also be seen to be struggling under the luminosity demands of GRB 051227, but this is almost certainly due to it having been placed at its redshift upper limit of $z = 2.8$. Indeed, when a higher efficiency is used (analogous to a lower $z$), this burst is well described by the model.

According to \citet{Metzger11} and \citet{Bucciantini12}, bursts with dipole fields $\gtrsim 10^{15}$ G will produce winds that are sufficiently clean to become optically thin at the jet energy dissipation radius on timescales suitable for EE. Our results for $B$ find good agreement with this threshold, especially since our value for dipole efficiency is somewhat tentative at 5\%, and could easily be increased, resulting in a further increase in $B$.

As can be seen in Figure~\ref{fits}, most (if not all) burst light curves are type II. This is certainly the most likely of the 4 types identified in section 3, as the template light curves returned a `classic' type \textbf{$37$\%} of the time, but these synthetic curves suggest we should have roughly 3 type I, 4 type II and 1 each from types III and IV from our sample of 9 (although these are small number statistics). From the best fit curves, GRB 061210 and GRB 071227 could be considered candidates for the type I population, which just leaves an absence of `sloped' and `stuttering' bursts. The reason for this could well be that these classes are not readily identified as EE. As previously mentioned in section $3.3$, a type III `sloped' burst could easily be identified as a LGRB or SGRB due to the single component look given to the light curve when propellering and dipole emission produce similar luminosities. Similarly, the type IV `stuttering' bursts could be mistaken for a SGRB with a flare, or a LGRB with $T_{90} \approx 10$ s. These rarer classes could then simply be absent from the accepted EE population, whilst the type I and II bursts, which are indistinguishable when given the right prompt emission or data availability, constitute the entire EE category. This could have a knock-on effect in $M_d$; the derived values for $M_d$ are typically quite low (a few $10^{-3}$ $M_{\odot}$), but the missing classes are those that typically exhibit the most massive discs, skewing the mean values towards the lower end. However, the predicted paucity of type III and IV propellers means that this effect may not be particularly large.

Whilst the results using the exponential accretion rate enjoy a reasonable degree of success, the two power-law accretion rates appear rather less suited to the task. In all cases, the obtained best fits were of lesser quality than those found with an exponential decay, and the fits were frequently unable to model both emission components simultaneously, instead settling for the EE alone. In both power-law cases, fitting the steep decays after the cessation of EE meant that the dipole emission was also forced to drop off rapidly, plummeting to a level far below that of the plateau in the data. There was no significant difference between the $t^{-4/3}$ and $t^{-5/3}$ profiles. From this, it seems that an exponential accretion rate may be required for magnetic propellering to be a viable mechanism in EE GRBs. Another key requirement for a successful propeller is that the conversion efficiency of kinetic energy to EM radiation for propellered material needs to be fairly high ($\gtrsim 10\%$). It is believed that the efficiency of the highly relativistic prompt emission can be $\gtrsim 50\%$ \citep{Nousek06}, so an efficiency in the region of 10\% -- 40\% is not entirely unreasonable for the slightly softer EE, but it is uncertain whether this level of efficiency could be maintained over the entire extended tail.

\begin{figure}
\begin{center}
\includegraphics[width=8cm,angle=-90]{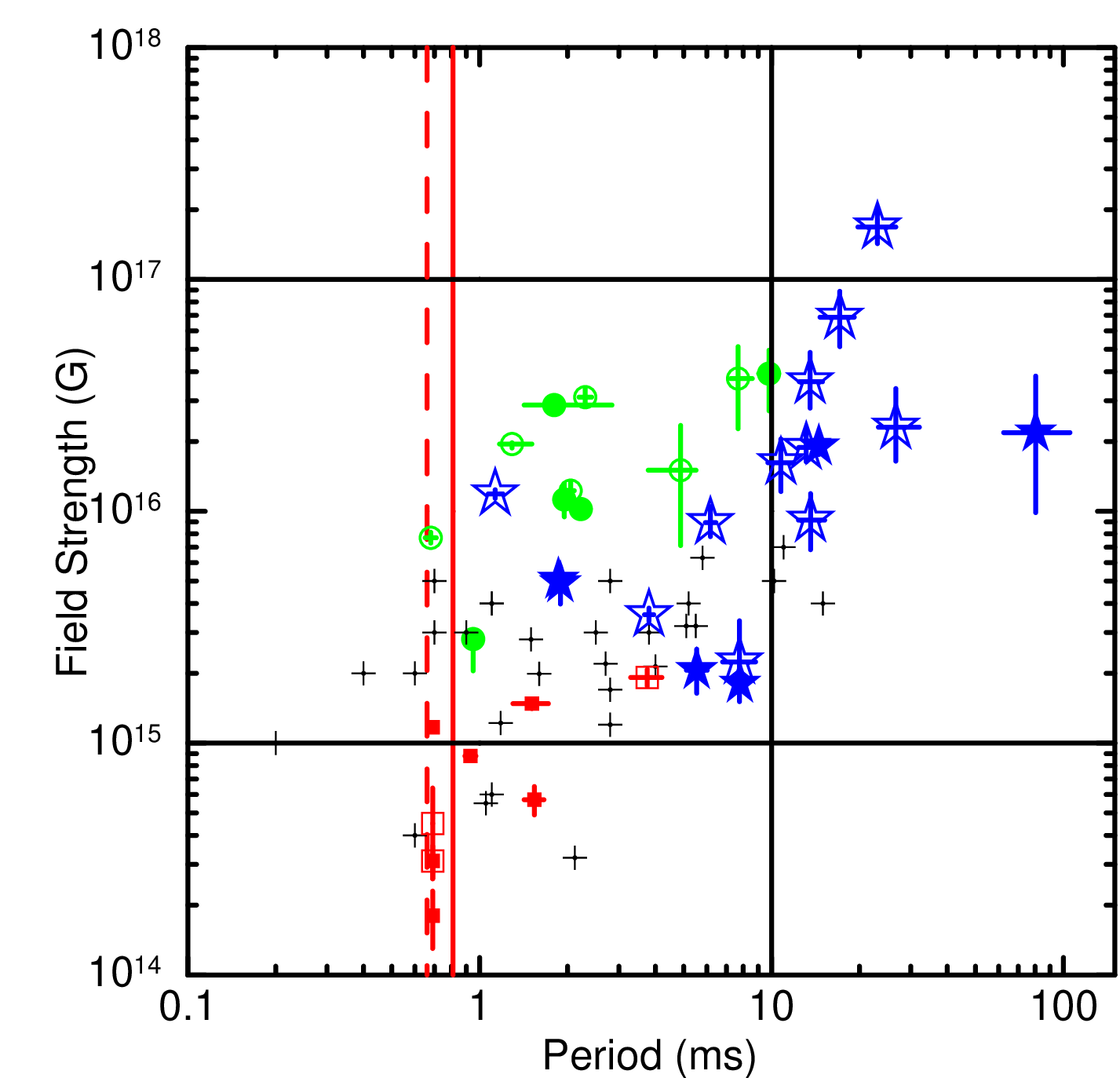}
\end{center}
\caption{A plot of magnetic field strength versus spin period. The solid (dashed) red line represents the spin break up period for a collapsar (binary merger) progenitor \citep{Lattimer04}. Blue stars: stable magnetars and green circles: unstable magnetars which collapse to form a BH \citep{Rowlinson13}. Black `+' symbols are the LGRB candidates identified by \citet{Lyons10,Dall'Osso11,Bernardini12}. The red squares are the magnetic fields and spin periods of the present work. Filled symbols have observed redshifts, open symbols use the sample average redshift, which is $z = 0.39$ for extended bursts and $z = 0.72$ for the short bursts from \citet{Rowlinson13}.}
\label{parspace}
\end{figure}

Figure~\ref{parspace} shows where the results place these EE bursts relative to other GRBs, both short and long. Whilst they appear to populate their own region of low B-field and spin period, caution is required when drawing conclusions from this plot. Firstly, these results where obtained using efficiencies of 40\% in the propeller and 5\% in the dipole, whereas, for example, the short sample from \citet{Rowlinson13} were examined using 100\% efficient emission. Secondly, and probably more importantly, the rate of dipole spin-down is enhanced by the presence of the accretion disc in the current work, making a direct comparison with previous results difficult, since they did not have this enhanced rate. If the enhanced rate is not used, then the values found for $B$ and $P$ in EE bursts lie in the same region of parameter space as those for the SGRB sample. Even if their spin periods and dipole fields are not unique, the degree to which magnetic propellering influences the light curves offers a natural explanation for the difference between the two classes, since any propeller luminosity is predicated on the presence of an accretion disc; remove the disc and you're left with an ordinary SGRB. In fact, the disc does not need to be completely absent. If the disc mass is below around $10^{-6}$ $M_{\odot}$ it becomes difficult to produce propeller luminosity much above $10^{49}$ erg s$^{-1}$ as the accretion rate is too low. As a result, emission becomes dominated by the dipole contribution and light curves take on forms increasingly resembling SGRBs (e.g. \citealt{Rowlinson13}).

\begin{figure}
\begin{center}
\includegraphics[width=8cm,angle=-0]{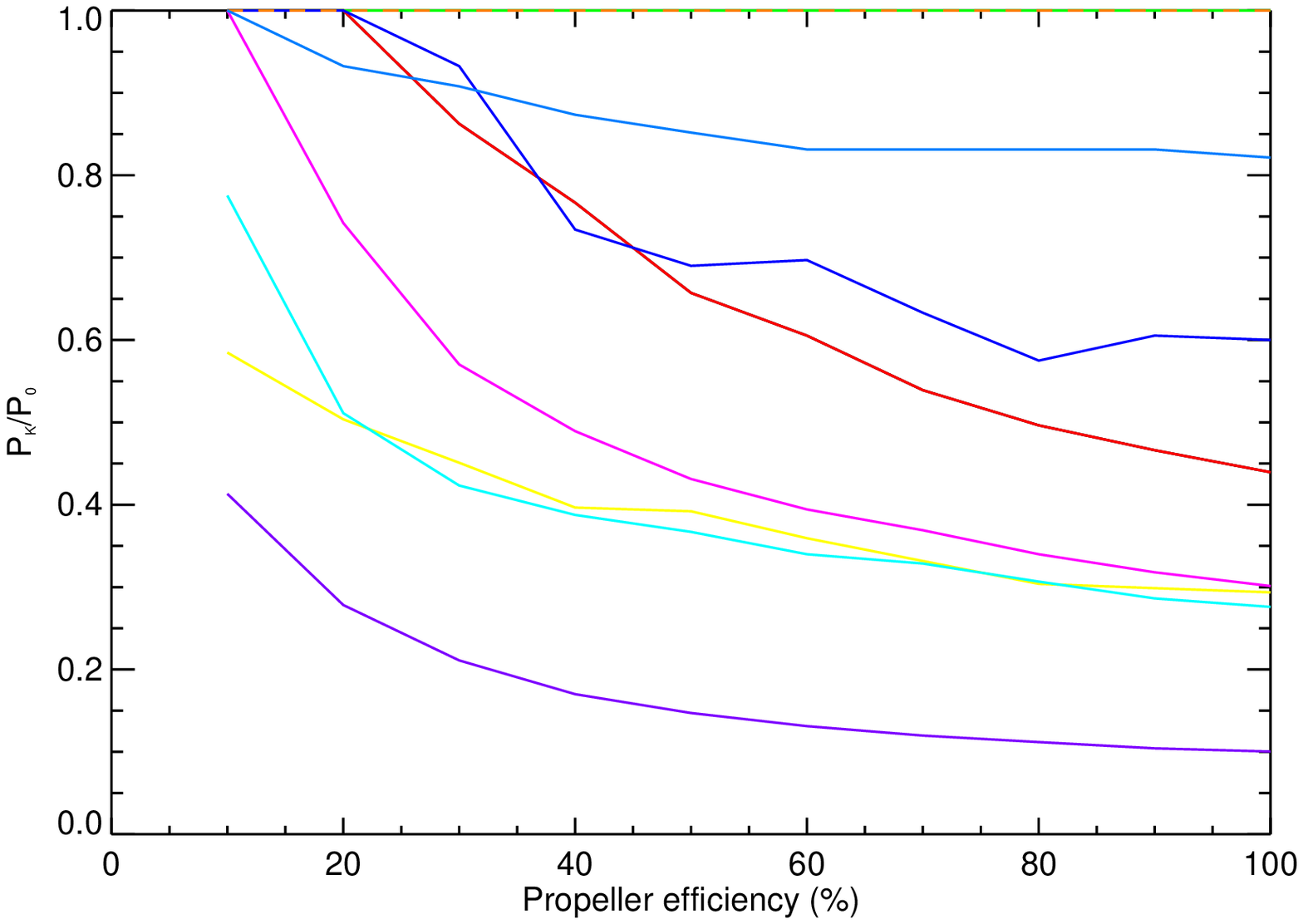}
\end{center}
\caption{Spin break-up period ($P_K$) over initial spin period ($P_0$) versus propeller emission efficiency. A value of 1 on the y-axis indicates the burst is born with $P_0 = P_K$, with decreasing fractions representing increasingly lower initial spin periods. Dipole efficiency is 10\% throughout. Red - 050724; Green - 051227; Blue - 060614; Yellow - 061006; Magneta - 061210; Orange - 070714B; Cyan - 071227; Light green - 080123; Violet - 111121A.}
\label{effplot}
\end{figure}

Creating discs of different masses requires varying conditions in the progenitor system. Two potential factors during binary merger are the mass ratio and the equation of state. \citet{Hotokezaka13} find the rest mass and kinetic energy of ejected material is greater with decreasing mass ratio (more asymmetric binaries) when the equation of state allows for more compact NSs. \citet{Rosswog07} also showed that binary systems with significantly unequal masses exhibit progressively more varied fallback behaviours with decreasing mass ratio. If material returns to the newly formed magnetar at earlier times and in greater quantities as described in \citet{Lee09}, then the conditions for propellering may be met. EE then, could be the product of an unequal mass binary merger, whilst SGRBs are born of more equal mass binaries. The comparitive rarity of EE events may be attributed to the lesser abundance of more massive ($\gtrsim 1.4$ $M_{\odot}$) NSs \citep{Valentim11,Lattimer12} and hence fewer unequal mass NS binaries.

\subsection{Radio emission}
The radio afterglow is one of the main proving grounds for the magnetar model. The presence (or lack) of radio emission on timescales of a few months to years after a burst is detected will place firm limits on the circumburst medium (CBM), or, in cases where the local density is already known, the magnetar model. Recently \citet{Metzger13} claim to have ruled out long-lived millisecond magnetars as the central engine for two bursts: GRB 060505 and GRB 050724, the latter of which features in this study. The authors found that a few $10^{52}$ erg ejected at $\beta_0 \sim 1$ should have been detectable during their observation $\sim 2.5$ years after the burst for $\epsilon_B = 0.1$, unless the CBM is $0.05$ cm$^{-3}$ or less. \citet{Panaitescu06} have independently constrained the CBM around GRB 050724 to be $0.1 < n < 10^3$ cm$^{-3}$, however \citet{Berger05b} find it to be consistent with values as low as $n \approx 0.02$ cm$^{-3}$, with a best fit value $n \approx 0.1$ cm$^{-3}$. The lack of detection could be explained if the value of $n$ lies at the lower end of this range. For higher densities ($n \gtrsim 0.05$ cm$^{-3}$) the lack of observation could be explained by a lower value of $\epsilon_B$. GRB 050724 is not typical even amongst the oddball sample of EE bursts, since it has the longest and one of the most luminous EE tails observed. In addition, it is unique in the class in having an as yet unexplained giant flare seen in the X-ray light curve at $\sim$ a few $10^4$ seconds after trigger.

For EE, ejecting the majority of a $10^{-3}$ $M_{\odot}$ fallback disc at initial velocities of up to $0.9c$ could produce a distinct feature in the radio signature of the GRB. We save the intricate details of this signature for future study, and discuss constraints on the model placed by previous radio band observations of EE bursts. Of the 9 EE GRBs in this sample, only GRB 050724 has a detection in radio emission, with 3 more (051227, 061210 and 070714B) having upper limits \citep{Chandra12}. All observation were taken using the VLA. Using the equations in \citet{Nakar11} (and supplementary information), we find the peak synchrotron frequency in the radio band to be more than an order of magnitude redder than the $8.46$ GHz observing frequency of the VLA. We also find the peak flux in the detector bandpass to be at least an order of magnitude lower than the afterglow detection in GRB 050724 \citep{Berger05b}, which was made during the late-time giant X-ray flare seen in the light curve at around $10^4$ -- $10^5$ s. The detected radio emission was fairly typical of SGRB radio afterglows, and we therefore believe it was the radio signature of the prompt emission, rather than that of the EE tail. The peak flux in the detector bandpass was also at least an order of magnitude lower than the upper limits for GRB 051227 \citep{Frail05}, GRB 061210 \citep{Chandra06} and GRB 070714B \citep{Chandra07}. We therefore conclude that radio observations of GRB afterglows are not currently constraining for EE if the underlying mechanism is a magnetic propeller, but are now at a level where they are becoming highly constraining to a millisecond pulsar (magnetar) central engine, and will become more so with the upgraded VLA \citep{Perley11}.

\section{Conclusions}
Using magnetic propellering and dipole spin-down, we have obtained the first simultaneous fits to both the extended tail and the afterglow plateau for a sample of 9 EE GRBs. We find typical disc masses of a few $10^{-3}$ $M_{\odot}$ to a few $10^{-2}$ $M_{\odot}$, spin periods of a few ms, and magnetic fields of around $10^{15}$ G. The ability to reconcile two emission features within a single central engine suggests there may be some weight to the idea that a highly magnetized neutron star is responsible for these phenomena. Whilst it is possible that the values for magnetic field and spin period are different in EE GRBs and SGRBs, it is hard to argue conclusively that this is the case. We suggest that the difference could also be due to subtleties in the progenitor, specifically the mass ratio, where unequal mass binaries produce the fallback material required to power magnetic propellering, whilst more equal mass binaries do not. We note that radio observations of EE GRBs are now at a level close to where magnetar spin-down can be ruled out, but do not appear to be constraining to the EE tail if the underlying mechanism is indeed a magnetic propeller. The major constraint currently is the requirement that the conversion of kinetic energy to EM radiation in accelerated material be at least $\gtrsim 10\%$.

\section{Acknowledgements}
BG acknowledges funding from the Science and Technology Funding Council. The work makes use of data supplied by the UK \emph{Swift} Science Data Centre at the University of Leicester and the \emph{Swift} satellite. \emph{Swift}, launched in November 2004, is a NASA mission in partnership with the Italian Space Agency and the UK Space Agency. \emph{Swift} is managed by NASA Goddard. Penn State University controls science and flight operations from the Mission Operations Center in University Park, Pennsylvania. Los Alamos National Laboratory provides gamma-ray imaging analysis. We thank the anonymous referee for helpful comments that improved the manuscript.

\bibliographystyle{mn2e}
\bibliography{ref}

\label{lastpage}

\end{document}